\documentclass[10pt,a4paper]{article} 

\usepackage[margin=1in]{geometry}
\usepackage{slashbox}

{}
{}
{}
\usepackage{url}
\usepackage{enumerate}
\usepackage{graphicx}
\usepackage{braket}
\usepackage{todonotes}
\usepackage{amsmath}
\usepackage{hyperref}
\usepackage{mathrsfs}
\usepackage{amsfonts}
\usepackage{mathtools}
\usepackage{comment}
\usepackage{longtable}
\usepackage{multicol}
\usepackage{multirow}

\usepackage{wrapfig}
\usepackage{lscape}
\usepackage{rotating}
\usepackage{makecell}
\usepackage{authblk}
\usepackage{subcaption}

\title{Synergy of machine learning with quantum computing and communication}

\author[1,*]{Debasmita Bhoumik}

\author[1,+]{Susmita Sur-Kolay}
\author[2]{ Latesh Kumar K J}

\author [2]{ Sundaraja
Sitharama Iyengar}

\affil[1]{Advanced Computing \& Microelectronics Unit, Indian Statistical Institute, Kolkata, India}

\affil[2]{KFSCIS, Florida International University, Miami, Florida,USA}

\affil[*]{debasmita.ria21@gmail.com}
\affil[+]{ssk@isical.ac.in}

\date{}

\begin{document}
\maketitle

\begin{abstract}
    Machine learning in quantum computing and communication provides intensive opportunities for revolutionizing the field of Physics, Mathematics, and Computer Science. There exists an aperture of understanding behind this interdisciplinary domain and a lack of core understanding renders an opportunity to explore the machine learning techniques for this domain. This paper gives a comprehensive review of state-of-the-art approaches in quantum computing and quantum communication in the context of Artificial Intelligence and machine learning models. The paper reviews the classical ML models that have been employed in various  ways for quantum computation such as quantum error correction, quantum communication, quantum cryptography, and mapping quantum algorithms to the existing hardware. The paper also illustrates how the relevant current challenges  can be transformed into future research avenues.
\end{abstract}

\section{Introduction}

One of the most notable events occurring in the fields of Computer Science and Physics is quantum computation. It can provide a faster solution to some of the real-life problems such as factoring a large integer \cite{Shor:1997:PAP:264393.264406}, searching a key in a large unsorted database \cite{Grover:1996:FQM:237814.237866}, simulating a Hamiltonian of a complex system \cite{childs2012hamiltonian}. With $N$ quantum bits (qubits), a quantum computer represents $2^N$ states. This exponential increase in the state space is a promising potential for many optimization problems which quickly become intractable even in modern supercomputers. 

About three decades ago, the quantum computation was primarily of interest to theoretical physicists and computer scientists. However, this field has made rapid progress in last few years. Both theory and practice have rapidly advanced in quantum computing and the interesting seminal area is how quantum computing may influence the domain of machine learning. The capability of quantum computers over classical computers was demonstrated by Google  \cite{arute2019quantum} where their $53$-qubit quantum computer was shown to execute random operations on a circuit within a few seconds which modern supercomputers would have required years. IBM showed the reliable execution of the largest quantum circuit till date - which is a trotterized circuit corresponding to  a nearest-neighbour Ising model \cite{kim2023evidence} with 127 qubits and a 2 qubit depth of 60. In the past few years, IBM has moved on  \cite{ibmresearchblog_2020} to a $433$-qubit one with an announcement for the launch of their new quantum device with 1,121 qubits in 2023. 

It was only a matter of time for quantum computation to cross paths with another domain of utmost interest and application in modern times: machine learning (ML), where each benefits from the other. ML essentially trains a computer to learn information from data without explicit programming. ML can be broadly classified into three categories: (i) \emph{supervised machine learning} (training data to predict output) such as Support Vector Machine (SVM), Neural Network (NN), Naive Bayes etc; (ii) \emph{unsupervised machine learning} (acting on unlabelled data) such as k-means clustering algorithm, and (iii) \emph{reinforcement learning} (learning from environment and mistakes through agent). Few popular examples of applications of classical ML are self-driven cars, fraud detection, natural language processing, online recommendation (from Amazon, Walmart, Hulu, Netflix).

In recent literature, we note that classical ML models have been applied in different portions of quantum computation such as quantum error correction, quantum communication, quantum cryptography, mapping quantum algorithms to the existing hardware etc. On the other hand, quantum counterparts of existing ML algorithms have been designed, such as Quantum Neural Network (QNN), Quantum Support Vector machine (QSVM) which are expected to perform better than the classical computers due to the exponential increase in the search space. Moreover, near-term quantum-classical hybrid algorithms have been designed which, in some sense, mimic the working principle of ML algorithms. They searched for the global optimum for a cost function, but uses quantum computers for the same.  There is already an interesting survey paper about how machine learning meets quantum foundations (mathematical and conceptual understanding of quantum theory) \cite{bharti2020machine}. In \cite{HOUSSEIN2022116512}, \cite{zeguendry2023quantum}, \cite{cerezo2022challenges} and \cite{garcia2022systematic}, the authors provided a detailed review of recent advancement in quantum machine learning. In this article,  we mainly focus on  how machine learning meets quantum computation ( branch of computer science is based on the principles of the quantum mechanics).

We briefly review each of the topics mentioned above. This is not an in-depth review of quantum computation and machine learning. Rather, the aim is to provide the readers with a broad overview of the various directions in which these two domains have interlaced. We enlist relevant references for the readers to delve deeper in some of the directions covered in this article, which does not assume prior knowledge of either quantum computation or machine learning. 

Section \ref{introtoqc} introduces  quantum computation briefly and Section \ref{introtoml} presents the basic concepts of machine learning. In Section \ref{ml_in_qc} we discuss how classical machine learning can be applied effectively for efficient design of quantum computing circuits, particularly  logic synthesis, physical mapping, and error decoding. Section \ref{nisq} addresses the techniques adopted for the present-day technology having small noisy quantum devices, such as variational algorithms and quantum approximate optimization algorithms (QAOA). In Section \ref{qml}, we discuss the key aspects of quantum machine learning. Next, Section \ref{qcomm} outlines the essence of quantum communication systems, quantum cryptography protocols and application of classical machine learning to design these. The concluding remarks appear in Section \ref{con}.

\subsection{A layman's introduction to quantum computation}
\label{introtoqc}
In classical computers, the basic unit of information is a bit (value = 0 or 1), whereas a quantum bit (or qubit) is its counterpart in quantum computing. It is represented as a unit vector in a 2-dimensional Hilbert space which is a complex vector space with two orthogonal basis states. These two are represented as $\ket{0} = (1 \quad 0)^T$ and  $\ket{1} = (0 \quad 1)^T$. The state of a qubit may also be any linear \emph{superposition} of  the two basis states. Thus, the state of a qubit can be of the form $\ket{\psi} = \alpha\ket{0} + \beta\ket{1}$, where $\alpha, \beta \in \mathbb{C}$ are termed as probability amplitudes and  $|\alpha|^2 + |\beta|^2 = 1$.

A \emph{quantum computer} employs operations which involve \emph{superposition},  \emph{entanglement} and \emph{interference}, which are not observed in the macroscopic classical domain.
\begin{itemize}
    \item Superposition is a quantum phenomenon with no classical equivalence. Let us consider the analogy of the act of flipping a coin. While the coin in mid-air is in both head and tail state simultaneously, which can be considered as a superposition, it consistently {\it lands} on either head or tail but not both - a binary phenomenon. Similarly, the quantum superposition for a state is lost upon measurement as the outcome is exactly one of the two basis states. Therefore, any operation on a superposition state can be considered as a simultaneous operation on all the basis states that form the superposition. This property leads to speedup in quantum algorithms.
    
    \item  Entanglement is another interesting phenomenon in quantum systems. Two qubits are said to be entangled if the measurement of one qubit disturbs the state of the other, irrespective of the spatial distance between them \cite{einstein1935can}. This property is valuable for quantum cryptography.
    
    \item  Interference is usually observed in waves. But since quantum states have wave-particle duality, i.e., a qubit can behave both as a wave and particle at the same time, two qubits can interact either constructively or destructively. When two qubits interact constructively, the corresponding probability amplitudes increase beyond the normal additive value. On the other hand, when they interact destructively, the probability amplitudes cancel each other out \cite{griffiths1960introduction}. The design of a quantum algorithm exploits this phenomena. The states corresponding to the correct solution(s) are made to interact constructively, whereas the other outcomes are made to interact destructively.
\end{itemize}

A brief summary of operations (gates) and the notion of a quantum circuit is given in Appendix.

Since the early days of quantum algorithms, two quantum algorithms which demonstrated the superiority over their classical counterparts, have been  of utmost interest due to their applicability. Grover's Search \cite{Grover:1996:FQM:237814.237866} exploits interference to  search for an element (called the marked state) in an unordered database faster. In particular, a classical computer would require $\mathcal{O}(N)$ queries to search for a marked state among $N$ states, whereas Grover's search on a quantum computer suffices with $\mathcal{O}(\sqrt{N})$ queries to provide a quadratic speedup. For example, in order to find an item in a list of one trillion items, where checking each item requires one microsecond, a classical computer would take approximately a week, but a quantum computer only about a second.

The second quantum algorithm  by Shor \cite{Shor:1997:PAP:264393.264406} that toasts the triumph of quantum computing is for factorizing an integer. Till now, there is no known classical algorithm that can achieve this task in polynomial time. The security of cryptosystems such as RSA relies on the classical hardness of factorization. Shor's algorithm thus renders such classical cryptosystems vulnerable in a quantum world.

These two and many other ones designed later on \cite{zoo}, have intensified the interest in quantum computing and the impetus to build required hardware.

\subsection{Machine learning algorithms relevant to quantum computation} 
\label{introtoml}
For the sake of completeness, before discussing the amalgamation of quantum computing and machine learning, we briefly discuss a few machine learning (ML) algorithms which are used extensively in the domain of quantum computation. As stated before, the goal of  machine learning  is to train a computer  to learn certain properties from a given dataset without explicit coding or a set of rules, and then use the outcome to study those properties in new data for prediction or classification purposes. For example, we expect that a computer that has previously seen a large number of pictures of tumors, and has been trained on the malignant ones,  would be able to identify an unseen photo of a tumor correctly. This type of ML algorithm is called supervised learning, where the machine is previously trained with some labeled data. Other forms of learning such as unsupervised and reinforced have also been studied widely and applied to various domains.

Three of the most widely used supervised machine learning algorithms related to quantum computing are (a) Neural Network (NN) \cite{mcculloch1943logical} for quantum logic synthesis, physical mapping, and quantum error decoding in Section \ref{qcs}), QKD protocol in Section \ref{qkd}, quantum accelerator of ML in Section \ref{aspdac}, the quantum neural network in Section \ref{qnn}, (b) Reinforcement learning (RL) \cite{sutton2018reinforcement}  for quantum error decoding in Section \ref{qcs}, and (c) Support Vector Machine (SVM) \cite{vapnik2013nature} for quantum machine learning in Section \ref{qsvm}. The study also discusses various ML learning models including the random forest method on quantum communication in Section \ref{qcs}.

\subsubsection{Neural Network}

The human brain contains $200$ billion neurons and each neuron consists of four parts: dendrites, soma, axon, and synapses. Signals are collected by neurons through dendrites, and then all the signals collected are summed up by soma. After reaching the threshold, the signal is passed to the other neurons through the axon. The power of the connection between the neurons is indicated by the synapses.

\begin{figure}[htb]
     \centering
     \begin{subfigure}[b]{0.45\textwidth}
         \centering
         \includegraphics[scale=0.24] {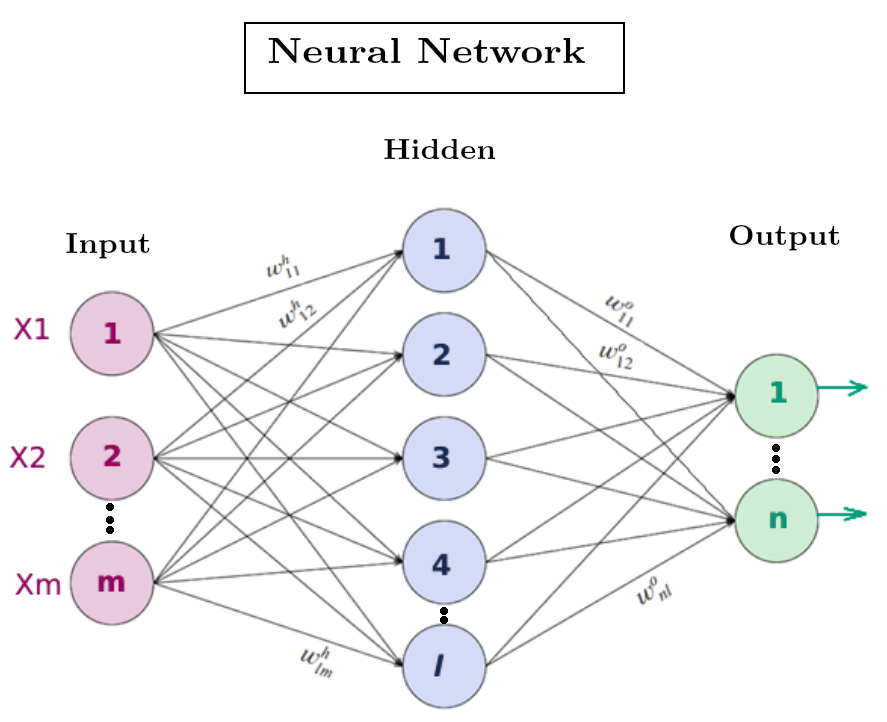}
         \caption{}
         \label{nn}
     \end{subfigure}
     \hfill
     \begin{subfigure}[b]{0.45\textwidth}
         \centering
         \includegraphics[scale=0.32] {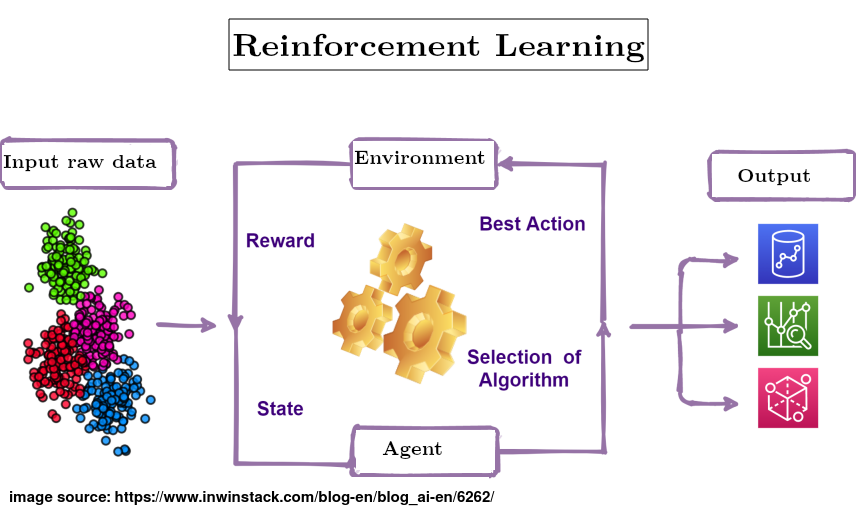}
         \caption{}
         \label{rl}
     \end{subfigure}

    \caption{ Schematic diagram of (a) Artifical  Neural Network [Image source: https://medium.com/swlh/neural-networks-4b6f719f9d75], and (b) Reinforcement Learning [Image source: https://www.inwinstack.com/blog-en/blog\_ai-en/6262/] }
    \label{fig:LRNN}
\end{figure}

Similarly, an Artificial Neural Network (ANN or sometimes called just NN) mimics this biological neural network. ANN was discovered in the year of 1943 by the neurophysiologists Warren McCulloch and the logician Walter Pitts  \cite{mcculloch1943logical}. In an ANN there are multiple layers representing the neurons. Simpler ANN algorithms have no feedback between the layers and are called feed-forward neural networks (FFNN). A single-layer FFNN consists of an input layer of neurons and an output layer of neurons. In a multi-layer feed-forward network, the first layer is the input layer which receives an input signal and the last layer is the output layer. In between these two layers, there can be multiple hidden layers. The signal from the input layer passes through these hidden layers to the output layer. The connection between a pair of nodes (neurons) in two adjacent layers has an associated weight, which indicates the connectivity strength between these. The input to a particular layer is multiplied by the weight to create the internal value, which is altered by a threshold value before feeding to an activation function to get the output of that layer. That output is passed on to the next layer as its input. The final layer provides the outcome of the network. In each iteration, the weights and the threshold values are updated to produce a more accurate value.  Figure \ref{fig:LRNN} (a) shows a neural network that has an input layer with $m$  nodes, one hidden layer with $l$ nodes, and an output layer with $n$ nodes. The inputs are $x_1, x_2, ... , x_m$, the weights on connections between the input layer and the  hidden layer are $w^h_{11}, w^h_{12}, ... , w^h_{lm}$. If the outputs can be directed back as inputs to the same layer or previous layers, then it results in a feedback neural network such as Recurrent Neural Network (RNN) \cite{medsker2001recurrent}.

\subsubsection{Reinforcement learning}
Reinforcement learning (RL) is a sub-field of machine learning that trains an agent to choose an action from the action space (where the environment is fixed), in order to maximize rewards over a specific time \cite{sutton2018reinforcement}. It is neither supervised nor unsupervised. Rather it receives a reward or penalty based on its choice of action. The algorithm learns to choose those actions which maximize its reward. There are four important elements in RL:
\begin{itemize}
    \item Agent -- a program which is trained to do a specific job;
    \item Environment --  the real or virtual world where the agent lies;
    \item Action -- a move which is made by the agent causing a change of status in the environment;
    \item Reward -- the evaluation after an action made by the agent and this may be either positive or negative.
\end{itemize}

While building an optimal strategy, an agent, can run into a dilemma between exploring new states and maximizing the overall reward at the same time. For example, it is impossible for the agent to know whether it has already reached a good enough reward and further exploration of new states will simply reduce its reward value. This is known as \textit{Exploration vs Exploitation}. The best overall strategy involves short term sacrifices to make the best overall decision. Reinforcement learning finds applications in multiple domains where training data may not be readily available.

\subsubsection{Support Vector Machines}
\begin{figure}[htb]
    \centering
   
    \includegraphics[height=5 cm, width= 12.5 cm] {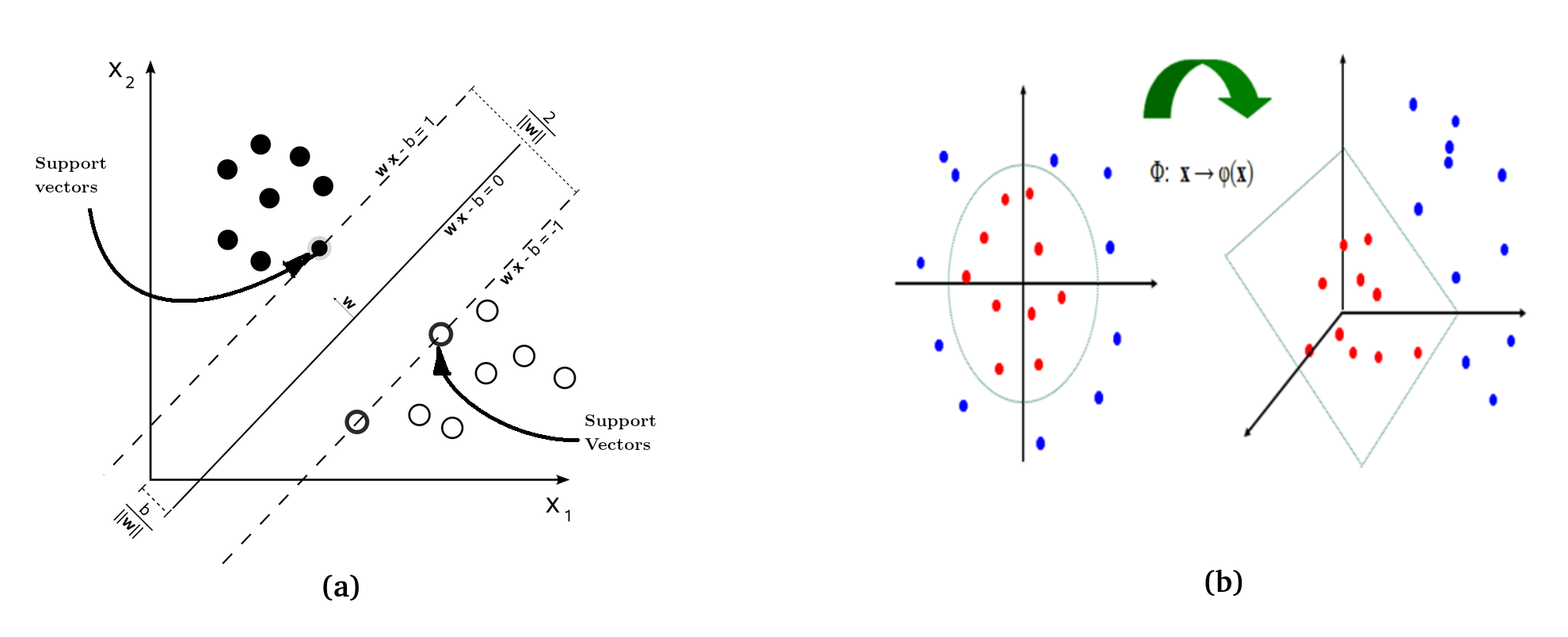}
    \caption{Support vector machine for classification. (a) Linear SVM. (b) Non-linear SVM and use of Kernel function to project the data in higher dimensions.  Here $\phi$ denotes a Kernel function}
    \label{fig:svm}
\end{figure}

Support Vector Machine (SVM) is a popular supervised ML algorithm. In this algorithm, the training set is of the form $(x_1, y_1)$, $(x_2, y_2)$, $\hdots$, $(x_n, y_n)$ where $x_i \in \mathbb{R}_d$, the $d$-dimensional feature space, and $y_i \in \{-1, +1\}$, the class label, with $i = 1 \hdots n$ \cite{vapnik2013nature}. An optimal separating hyperplane is built by the algorithm, based on a kernel function $K$. Depending on the feature vector, the data which lies on one side of the hyperplane belong to class $-1$, and the rest belong to class $+1$ (Fig. \ref{fig:svm}). If a straight line is unable to separate all the data points, then we need a nonlinear SVM classifier. It uses kernel functions ($\phi$) such as linear, polynomial, radial basis function (RBF), and sigmoid kernel. These functions project the data to a higher dimensional space so that they become linearly separable in that space. For multi-class SVM, two techniques have been used: (i) one-against-one which integrates several binary classifiers, and (ii) one-against-all which examines all the data at once.

In the next section, we study the application of various ML methods for different aspects of quantum computation.


\section{Classical machine learning in quantum computing}
\label{ml_in_qc}
Classical machine learning approaches find applications in different aspects of quantum computing such as logic synthesis, mapping a quantum algorithm to the underlying hardware, and decoding a quantum error correcting code. In the following subsections, we briefly touch upon each of these application domains, and show how machine learning leads to enhanced performance.

\subsection{ In efficient design of quantum computing systems}
\label{qcs}
A quantum circuit has a certain number of qubits and the algorithm is executed by a sequence of operations by quantum gates on thses qubits.
After designing a circuit, it has to be mapped to a quantum hardware (also termed as device which realizes the qubits as quantum systems and the gates by application of electromagnetic pulses). Now the device layout can constrain the qubit connections for two-qubit gates. This problem can be solved by layout synthesis. It produces an initial mapping in the quantum computing devices from the circuit qubits to the physical qubits. It adjusts the mapping by legalizing two-qubit gates with insertion of a chain of SWAP gates as needed to bring the two qubits under operation in close proximity for propoer gate operation. It schedules all the gates with a constraint that  the original functionality has to be invariant and is executable on the quantum computer with minimal quantum resources and execution time.

Mapping a quantum circuit on hardware typically consists of the following steps:
\begin{enumerate}
    \item \textit{Virtual circuit optimization}: In this step the input quantum circuit is optimized. This is obtained by applying various logic identities. For example, $HXH = Z$; so the three gates in a series can be replaced by a single gate only. Furthermore, if a gate $U$ is followed by $U^{\dagger}$, then those two can be replaced with identity.

    \item \textit{Decomposition of gates with 3 or more qubits} Most of current quantum hardware are designed to execute only one and two qubit gates. Therefore, all gates involving three or more qubits are decomposed into a cascade of single and two qubit gates.

    \item \textit{Placement}: In this step each virtual qubit from the quantum circuit is allocated to a physical qubit of the hardware.
    
   \item \textit{Routing}: The gates are scheduled for each qubit so that the depth of the circuit is minimized while maintaining their temporal ordering in the original circuit. This step takes the underlying hardware connectivity constraints into consideration and inserts necessary swap gates. Routing and  initial qubit allocation are interdependent and both optimization are computationally hard problems.
 
\item \textit{Translation to basis gates}: Each gate of the circuit are translated to the basis gate set of the underlying hardware. For IBM Quantum devices, the basis gate set is $\{X, SX, RZ, CX\}$.
    \item \textit{Physical Optimization}: Finally the rewritten or {\it transpiled or transformed}
circuit obtained so far is further optimized for resources and depth.

\end{enumerate}

\begin{figure}[htb]
    \centering
   
    \includegraphics[height= 6 cm, width= 13 cm] {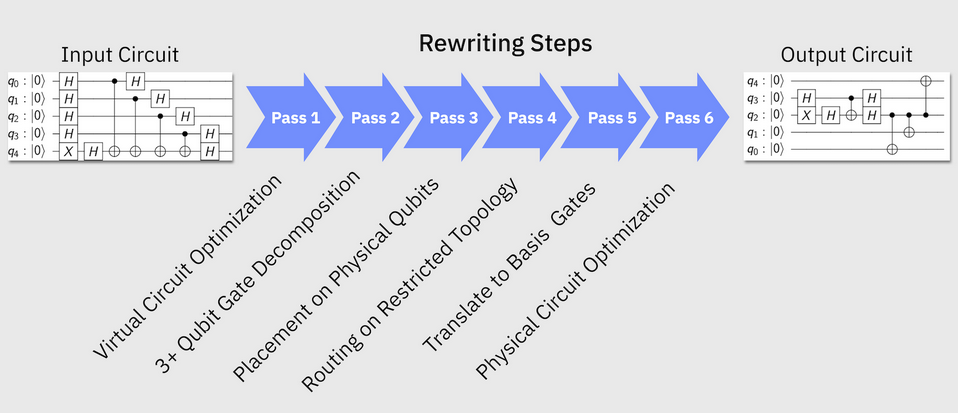}
    \caption{Mapping of quantum circuit on hardware by QisKit transpiler \cite{transpiler}}
    \label{fig:map}
\end{figure}

The flow of quantum circuit mapping is shown in Fig. \ref{fig:map} \cite{transpiler}. Most of the placement algorithms  resort to  heuristic methods \cite{bahreini2015minlp, cowtan2019qubit, wille2019mapping}, including ML \cite{paler2020machine}. With the increase in the size of quantum device to thousands of qubits in future for meaningful computation, it may be of importance to design ML algorithms that can accomplish this task within a reasonable time. This canemploy either  reinforcement learning or supervised learning, where the model is trained on smaller devices and is applied for placement on larger devices.

\subsubsection{Optimization in quantum circuit mapping}
In the early days, quantum circuits comprised of only a dozen of gates and upto 5 qubits. Recently,  experiments on circuits having 127 qubits and gate depth of 60 \cite{kim2023evidence}, with single qubit gates in the range of thousands, and two-qubit gates in the range of hundreds \cite{arute2019quantum}, as well as circuits with 433 qubits  have been carried out. The number of possible initial mappings for the former circuit will be $127!$ and the subsequent scheduling and legalization steps have a huge search space.

A circuit mapping method based on the $A^*$ algorithm was proposed which intends to find the best placements by comparing it to the desired outcome \cite{zulehner2018efficient}. A general drawback of this method is that the desired outcome is often unknown. 
The optimality of a layout method can be expressed in many ways, for example, by optimizing the depth of the circuits \cite{paler2020machine}. The QCL takes an input circuit $C_{in}$ to transform it to the functionally equivalent $C_{out}$ circuit. The  output circuit depth may be greater than the input circuit and the ratio between the depth of the output and the input circuit is the benchmark of the QCL methods. The minimization of the ratio is known as the QCL optimization. In \cite{paler2020machine} the authors have introduced a layout method  QXX,  including a  configurable  Gaussian  function for estimating  the  depth of  the generated circuits and to determine the circuit region that impacts most of the  depth.  The parameters of the QXX  model is optimized using  an improvised weighted random search. They introduced  QXX-MLP which is a multi layer perceptron neural network for predicting the depth of the circuit which is generated by QXX. After comparing QXX and QXX-MLP with the baseline they conclude that  QXX performs equally with  state-of-the-art  layout  methods \cite{tan2020optimality}.

\begin{figure}[htb]
     \centering
     \begin{subfigure}[b]{0.3\textwidth}
         \centering
         \includegraphics[scale=0.4] {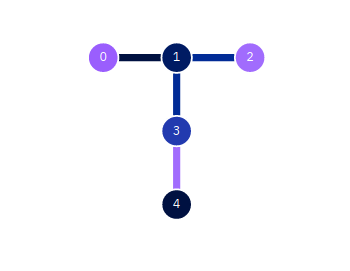}
         \caption{}
         \label{Quito}
     \end{subfigure}
     \hfill
     \begin{subfigure}[b]{0.65\textwidth}
         \centering
         \includegraphics[scale=0.25] {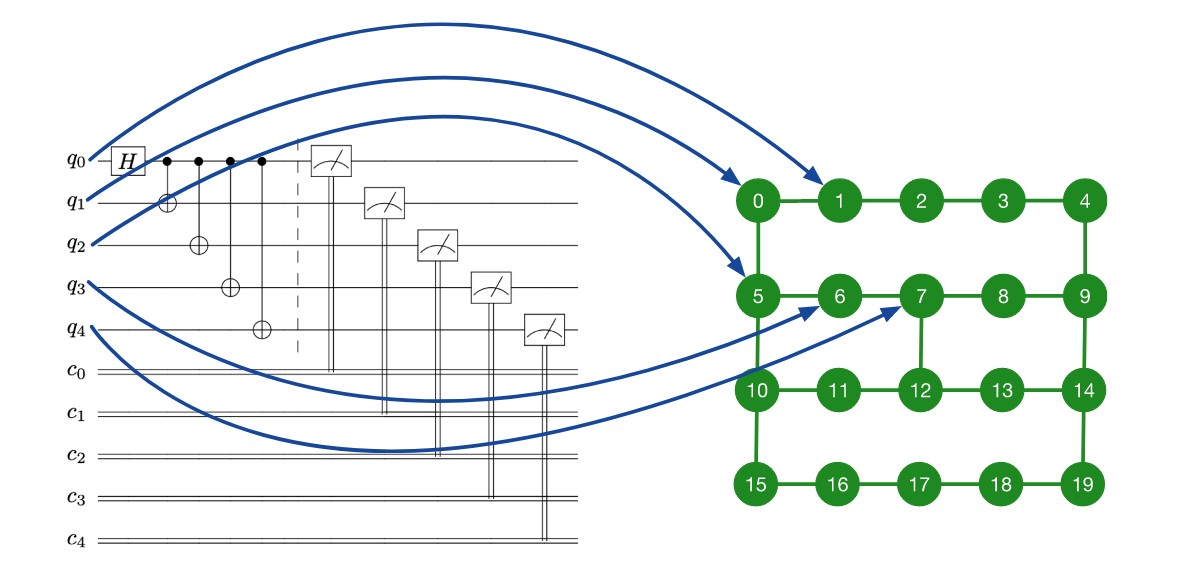}
         \caption{}
         \label{nn}
     \end{subfigure}

    \caption{(a) Coupling map of a 5 qubit IBM-quantum processor Quito, (b) graphical representation of a circuit mapping  \cite{acampora2021deep} where a 5-qubit circuit is mapped on a 20-qubit quantum processor with a given coupling map}
    \label{fig:mapping}
\end{figure}

There is a  gap between the quantum resources required for the execution of modern quantum algorithms and the resources available in current Noisy Intermediate--Scale Quantum (NISQ) devices,  because the processors consist of a sparsely interconnected coupling map, thereby limiting the interactions among the qubits. In Fig.\ref{fig:mapping}(a), we show the coupling map of the 5 qubit Quito IBM Q processor. This coupling map defines the set {(0,1),(1,0),(1,2),(2,1),(1,3),(3,1),(3,4),(4,3)} where the tuples are target and control respectively in a possible CNOT operation. Therefore, eight  of the twenty (i.e., $n^2 - n$) pairs can be used for CNOT operation in a circuit. This limits the  opportunities  offered  by quantum devices. Hence efficient circuit mapping techniques are needed for SWAP gate minimization to perform CNOT operations between other pairs of qubits.

In \cite{acampora2021deep} the authors have proposed  a method employing deep neural networks (DNN) for quantum circuit mapping to improve the performance of current heuristic cost based methods. Their method comprises three  steps:
\begin{itemize}
    \item  formulating the circuit mapping problem as a classification problem;
    \item training the DNN for the classification task of circuit mapping with appropriate dataset;
    \item applying fine-tuning to make correct  predictions amenable with some logical constraints which is used to characterize the quantum processors.
\end{itemize}

Their formulation of the circuit mapping problem \cite{acampora2021deep} is as follows. Let $C_n = \{ c_i \}_{i \in \mathbb{N}} $ be the set of all $n$-qubit quantum circuits belonging to the set  $Q_c =\{q_1^{c},q_2^{c},..., q_n^{c} \} $, $P$ be a $m$-qubit $(\geq n)$ NISQ processor $\in$  $Q_P =\{q_1,q_2,..., q_m \} $ with the corresponding coupling map is $M_P = \{(q_i, q_j)|q_i, q_j \in Q_P$ and $i \neq j \} $. The initial circuit mappings consist of a set of functions $F = \{f_k : C_n \rightarrow \wp (Q_C \times Q_P) \}_{k \subseteq \mathbb{N}} $ where each $f_k$ associates the qubits of each circuit $\in$ $C_n$ to the qubits of the processor $P$. The final solution of the optimization problem is a function $f^* \in F $ which is used to minimize the number of swap operations needed.  For example, consider a $n=5$-qubit circuit to be mapped on a processor composed of $m=20$-qubits and a coupling map given in Figure \ref{fig:mapping}(a). A mapping $f$ is shown in Figure \ref{fig:mapping}(b).

A classification problem is defined as the task of evaluating a class label $y \in Y = \{L_1, L_2,...L_Q\}$ for a $K$-dimensional input vector $\mathbf{x} \in X \subseteq R^K $. For a given $n$-qubit quantum  circuit $c \in C_n$, and a $m$ qubit quantum processor $P$  $\in$  $Q_P$ distinguished  by  a specific coupling map $M_P$,  the quantum circuit mapping problem can be framed as classification in the following manner. It is implemented by a function $\phi$ which  for an input vector $\mathbf{x} \in X \subseteq R^K $ having a set of features  for the circuit $c$ and the processor $P$ where the circuit is to be mapped and executed. The features estimate the array $\mathbf{y}$ composed of $m$ elements, and each element belongs to the mapping label set $Y= \{-1\} \cup \{1,2,...,n\} \in \mathbb{N} $. The function $\phi$ is learnt  with a training  set of $N$ points which consists of pairs of feature set and mapping label representing a mapping from  $c$ to $P$, by a DNN based method.  This DNN is composed of one input layer, few hidden layers, and one output layer. 

Along with minimizing the number of swap gates, the authors also considered the error rate and the latency of each two-qubit gate with the aim to minimize both. The data set size for training and testing the DNN is composed of  $\sim$ 40000 random  quantum circuits, which are operating on 5 qubits and characterized by at most 10 CNOT gates. From each circuit, the number of extracted features is 22. There is an output label attached to each instance of the data set.
 They have demonstrated experimentally that this DNN can speed up the state-of-the-art circuit mapping algorithms used by IBM Qiskit Transpiler \cite{transpiler} for performing circuit mapping operations on 5-qubit IBM quantum processors, although popular algorithms available in IBM Qiskit, such as Dense Layout and Noise Adaptive Layout, can produce the best circuit mapping, with fewer number of SWAP gates. The DNN based method outperforms other popular machine learning techniques as the overall accuracy value of the best  classifier, Random Forest  is  15\% lower than that for this.

\subsection{In decoding error syndrome of a quantum error correcting code (QECC)}
\label{qecc}
The fragility of the qubits is the problem in realising  a large-scale quantum computer. The suggested solution to this problem is quantum error correction (QEC). Like classical error correction, quantum error correction also has an encoding process where the information of a single qubit is distributed into more than one qubit followed by a decoding process that identifies the error and corrects the noise that is inserted in the quantum system. Classically, this is easily achieved by the simplest 3-bit repetition code, where the encoder maps bit $0 \rightarrow 000$ and $1 \rightarrow 111$. The encoded bit-strings $000$ and $111$ are termed as the logical code-words. If the message incurs a single bit-flip error during transmission, then the receiver may get $010$. Hence, the receiver can interpret (decode) that the original code word was $000$ via  majority voting. But if the code word is subject to more than one bit-flip errors, the majority voting  leads to an incorrect code word. If it consists of 3 bit flips, then $000$ becomes $111$ which is also a valid code word, thus rendering error detection to be impossible. The distance $d$ of a code is defined as the minimum number of errors that can change a valid code word to another valid one; hence it is 3  here. It can be proved that the relation between the distance $d$ and the number of correctable errors $e$ is $e=(d-1)/2$.  Decoding an error syndrome implies mapping it to the error so that appropriation correction can be applied to eliminate the error.

\subsection{Shor's QECC}
For quantum systems however, if $\ket{\psi}$ denotes a general qubit in superposition, encoding of the form $\ket{\psi \psi \psi}$ is prohibited due to the no-cloning theorem. Hence it is necessary to design other forms of encoding to distribute the information of a single qubit into multiple qubits to form a logical qubit and is less prone to error. But errors can still occur at the physical level. Therefore, detection of those errors, often called decoding, are necessary to eliminate these and keep the state error free. However, the encoded state cannot be measured directly to detect the presence of error without destroying the information content in it. Therefore, extra qubits, called ancilla qubits, are required. They do not play a role in the actual encoding and computation, but are required to store the error information, also called syndrome, to be obtained via measurement and then decoded. 

An error on a qubit may be represented as an operator acting on it. In Shor \cite{PhysRevA.52.R2493}, it has been proven that a quantum error which can be expressed as a unitary operator, is a linear combination of the Pauli matrices, defined by the two-dimensional matrices given by: 
\begin{center}
    $I = \begin{pmatrix}
    1 & 0 \\ 0 & 1
    \end{pmatrix} $ \qquad $X = \begin{pmatrix}
    0 & 1 \\ 1 & 0
    \end{pmatrix} $ \qquad     $Z= \begin{pmatrix}
    1 & 0 \\ 0 & -1
    \end{pmatrix} $ \qquad      $Y = i \cdot Z \cdot X$
    
\end{center}
If a quantum error correcting code (QECC) can correct the Pauli errors, then  any unitary error on the system can be corrected by it.
A bit-flip error on a qubit or a Pauli $X$ error is given by $X \ket{0}= \ket{1}$, $X \ket{1}= \ket{0}$. Simialrly, a phase-flip on a qubit or a Pauli $Z$ error maps $Z\ket{0}=\ket{0}$,  $Z\ket{1}=-\ket{1}$. Creation of a logical qubit can be explained with a simple example. Let us consider a three-qubit code designed to detect a single-bit flip error. \\
Let the quantum state be $| \psi>  = \alpha |0> + \beta |1>$ be encoded as
$$\ket{\psi} = \alpha\ket{0} + \beta\ket{1} \xrightarrow{three-qubit ~encoder} \ket{\psi}_L = \alpha\ket{000} + \beta\ket{111} =\alpha\ket{0}_L + \beta\ket{1}_L.$$

\begin{figure}[htb]
    \centering
   
    \includegraphics[height=6 cm, width= 10 cm] {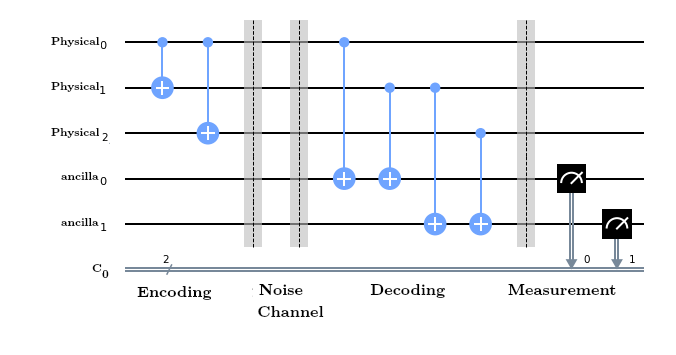}
    \caption{Encoding and decoding circuit for error correction : using Qiskit \cite{QiskitTextbook}, where the bit-flip error is acting on the first qubit }
    \label{fig:decoding}
\end{figure}

Therefore, $\ket{0}_L =\ket{000} and \ket{1}_L= \ket{111}$. This does not violate the no-cloning-theorem as $\ket{\psi}_L = \alpha\ket{000} + \beta\ket{111} \neq  \ket{\psi} \otimes \ket{\psi}$. Suppose there is a bit-flip error on the first physical qubit of 
the logical qubit which gives the state $X_1\ket{\psi}_L = \alpha\ket{100} + \beta\ket{011}$ , where $X_1$ is a bit-flip error on the first qubit. In the circuit of Figure \ref{fig:decoding}, the first part is for encoding, then there is an error channel, followed by  the decoding part and finally the measurement of the ancilla. 
 
In the decoding circuit, the two ancilla qubits are the target qubits of the CNOT gates. Thus, $\alpha \ket{100} \ket{00} + \beta \ket{011} \ket{00}$
$ \xrightarrow{CNOT3\&4} \alpha \ket{100} \ket{10} + \beta \ket{011} \ket{10} $
$ \xrightarrow{CNOT5\&6} \alpha \ket{100} \ket{10} + \beta \ket{011} \ket{10} $ =
$(\alpha \ket{100} + \beta \ket{011}) \ket{10}$.

From Table \ref{table:syndromedecoding}, we can see that the decoder can predict the location of the error by observing the ancilla outcomes .

\begin{table}[h!]
\centering
\caption{Error detection after measuring the two ancilla qubits}
\begin{tabular}{ |c|c| }
 \hline
Ancilla& Location of bit-flip error\\
 \hline
00 & No error\\
01 & $Physical_2$\\
10 &  $Physical_0$\\
11 & $Physical_1$\\
 \hline

\end{tabular}

\label{table:syndromedecoding}
\end{table}

For correction, we can simply apply the bitflip again at the detected position.  Similar to the classical coding, the distance $d$ of a quantum code is given by the minimum Hamming distance between two logical qubits \cite{roffe2019quantum}. A  logical Pauli operator   transforms a code-word state to another one. Shor first proposed the 9-qubit code \cite{PhysRevA.52.R2493} that corrects a single unitary error. Later on, a 7-qubit QECC by Steane \cite{PhysRevLett.77.793} and a 5-qubit QECC by Laflamme \cite{PhysRevLett.77.198} were proposed and the latter was shown to be optimal in the number of qubits for correcting a single unitary error.

\subsection{Topological QECCs} 
\begin{figure}[htb]
     \centering
     \begin{subfigure}[b]{0.3\textwidth}
         \centering
         \includegraphics[scale=0.2] {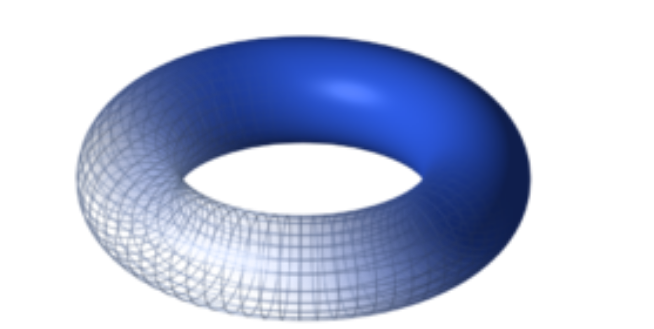}
         \caption{ }
         \label{torus}
     \end{subfigure}
     \hfill
     \begin{subfigure}[b]{0.6\textwidth}
         \centering
         \includegraphics[scale=0.5] {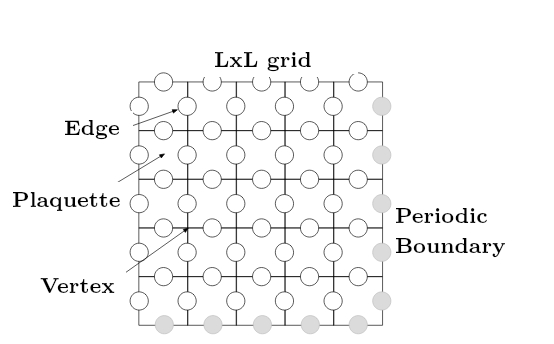}
         \caption{}
         \label{toric}
     \end{subfigure}

    \caption{ (a) A torus and a schematic lattice on it, (b) Lattice with $L=5$ for a Toric code }
    \label{fig:torusandtoric}
\end{figure}

The design of the encoder and decoder circuits for these QECCs often involve operations between non-adjacent qubits. This is costly as it requires one or more swaps among the neighbouring qubits, thus making the process slower and more error prone. This is called the Nearest Neighbour (NN) problem. In order to solve this drawback, topological code \cite{kitaev2003fault,bravyi1998quantum} came into place. The simplest topological code is a toric code, where the qubits are placed in a square lattice on the surface of a torus (Figure \ref{fig:torusandtoric} (a)) having periodic boundaries.

Consider an $L \times L$ square lattice on the surface of a torus, consisting of edges, vertices (points where edges meet) and plaquettes (individual square tiles enclosed by a set of edges  or 4-cycle faces). A qubit is associated with every edge on the lattice (indicated by circles on Figure \ref{fig:torusandtoric} (b). On an  $L \times L$ lattice with periodic boundary conditions  (the right-most edge is wrapped around and identified with the left-most edge and the upper edge with the lower edge), there are $2L^2$ edges.

\begin{figure}[htb]
     \centering
     \begin{subfigure}[b]{0.2\textwidth}
         \centering
         \includegraphics[scale=0.35] {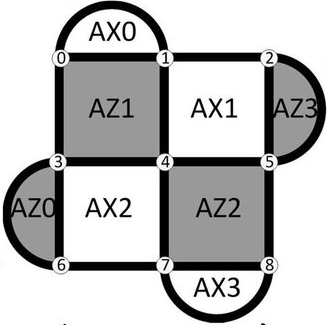}
         \caption{}
         \label{sc17}
     \end{subfigure}
     \hfill
     \begin{subfigure}[b]{0.7\textwidth}
         \centering
         \includegraphics[scale=0.35] {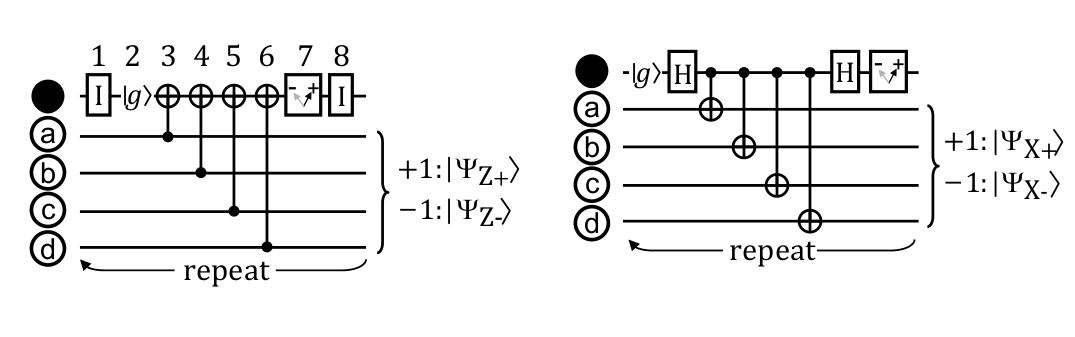}
         \caption{}
         \label{sc17ckt}
     \end{subfigure}

    \caption{Surface code: (a) Distance 3 rotated surface code \cite{varsamopoulos2017decoding}, (b) quantum circuit for one surface code cycle for a measure-Z and measure-X qubit \cite{fowler2012towards} }
    \label{fig:Surfacecode}
\end{figure}

Later, the toroidal structure developed by Kitaev was simplified to a planar version by Bravyi and Kitaev \cite{bravyi1998quantum} , and by Freedman and Meyer \cite{freedman2001projective}. This gave us {\bf surface codes}.
A schematic diagram of  surface code \cite{fowler2012towards, dennis2002topological,fowler2009high,wang2011surface,wootton2012high} is shown in Figure \ref{fig:Surfacecode}a.

 But with surface codes, the mapping from syndromes to error configurations is not one-to-one. Among the  syndrome decoding approaches union-find \cite{delfosse2021almost}, belief propagation \cite{fuentes2021degeneracy}, tensor network \cite{bravyi2014homological} and the minimum weight perfect matching (MWPM) \cite{edmonds1965paths, fowler2012towards, fowler2013minimum, chamberland2020topological} are used widely. For topological code, MWPM  achieves good decoding  performance, which is assessed in terms of its {\it threshold}, which is the physical error probability beyond which increasing the distance of the code leads to poorer accuracy \cite{fowler2012towards}. For MWPM, this threshold for surface code is  $\sim 6 \times 10^{-3}$ (for logical -X errors).
 
 \subsubsection{ML based decoding of surface codes} The worst case time complexity of MWPM scales as $O(N^3\log N)$ where $N$ is  the  number  of  physical qubits (i.e., $O(d^2)$, $d$ being the distance of the QECC) for a logical qubit. For example, $N$=$9$ when $d$=$3$.  A faster alternative is to apply machine learning techniques for  identifying  errors, as the decoding time scales linearly with the number of qubits \cite{varsamopoulos2019comparing} after the ML model has been trained and validated. These provide at least asymptotically similar decoding  performance  to  traditional  decoding  algorithms \cite{baireuther2018machine, krastanov2017deep, varsamopoulos2019comparing, chamberland2020topological, ni2020neural, sweke2018reinforcement, bhoumik2021efficient}. In \cite{varsamopoulos2019comparing} the threshold for surface code was  $\sim 3.2x10^{-3}$ using Recurrent Neural Network (RNN).
 Hence  Recurrent Neural Network shows a better trade-off between decoding performance and execution time.
 
 For depolarizing noise model, where the gates $X$, $Y$ or $Z$ apply to the data bits with a probability of $p/3$, and feed forward neural network based decoder, the threshold of the rotated surface code was reported as 0.146 whereas that of the Blossom (algorithm for MWPM) decoder is 0.142; so the ML model performs better for this noise model also.
 
There are a number of machine learning based decoders and the major differences lie in the structure of the ML model and its training algoirthm, and their decoding  performance  and  execution time differs  depending on the size of the training data set available and other related aspects. In \cite{torlai2017neural}, the authors present a decoding algorithm suitable for topological code where the decoder employs the simplest type of stochastic neural network to learn the structures which can make the approximate decoding problem easier than the general NP-hard decoding problem. They employ the restricted Boltzmann machine \cite{hinton2012practical} for unsupervised learning and test the  decoder numerically on a simple two-dimensional surface code with phase-flip  errors.  Given  an  error  chain $e0$ with  syndrome $S_0$, the  Boltzmann  machine  generates an error chain compatible with $S_0$ which can be used for the recovery. For this purpose, the network is trained  on different  datasets  obtained  for various values of $p_{err}$, the   probability of error. Their Boltzmann machine based neural decoder has achieved the error threshold of $0.109$. 

In \cite{varsamopoulos2019comparing}, the  error decoder has two levels. The aim of the low level decoder (LLD) is  to correct the errors in the individual physical data qubits, which makes the process extremely granular owing to the large number of data and measurement qubits. The high level decoder (HLD) tries to fix the errors in the logical qubit as a whole without considering the errors in the individual physical qubits. This eases the training processes for neural networks and thus provides better results. For the high-level decoder, a neural network and a non-neural network based {\it simple} decoder run in parallel to make an accurate prediction of the error correction required to rectify the errors due to noise. The authors considered two kinds of errors for creating the dataset to train and test the decoders on. First is the depolarizing noise and second is the circuit noise where both the gates and measurements are considered to be noisy. The results demonstrated in the paper suggest that the HLDss perform better than the LLD decoders. Furthermore, the RNN based HLDs perform better than the feed forward NNs HLDs in terms of decoding accuracy, but are slower than the FFNNs due to a larger number of parameters. The authors finally concluded that the RNN based HLDs create the best balance between decoding accuracy and prediction time for moderate variances in error probability.

Few papers have introduced reinforcement learning (RL) framework for topological codes  \cite{nautrup2019optimizing,DOMINGOCOLOMER2020126353,sweke2020reinforcement}. In \cite{nautrup2019optimizing}, the goal of the RL is to optimize and fault-tolerantly adapt  surface code. In \cite{DOMINGOCOLOMER2020126353}, the authors apply deep RL techniques to design decoders with high threshold for the toric code, which can operate under uncorrelated noise. They have found near-optimal performance around the theoretically optimal threshold of $0.11$.  

\subsubsection{ML based decoding of heavy hexagonal codes} 
\begin{figure}[htb]
    \centering
    \includegraphics[scale = 0.5] {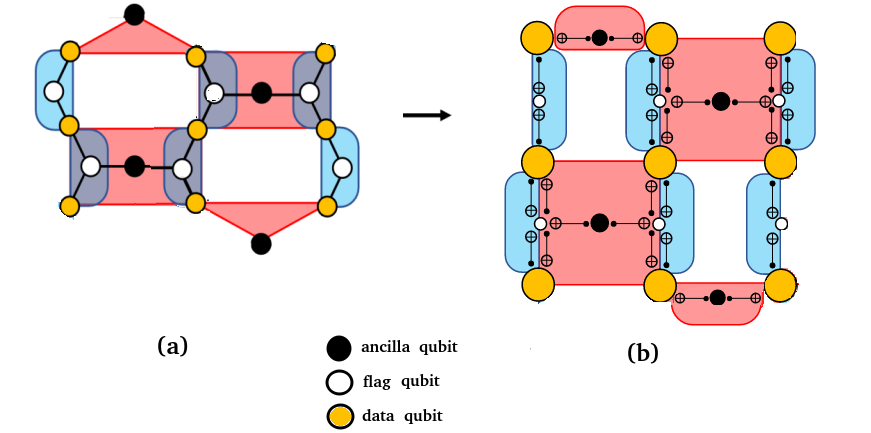}
    \caption{Distance 3 heavy hexagonal code encoding one logical qubit: (a) the hexagonal structure, (b) the circuit illustration of the heavy hexagonal code with the CNOT gates.
    Here yellow, white and black circles represents  data,
    flag and ancilla qubits respectively;  black ancilla qubits are for measuring the $X$ (red face or plaquette) and $Z$ (blue face or strip)  gauge generators. The product of two $Z$  gauge generators at each white plaquette forms a $Z$ stabilizer \cite{chamberland2020topological}. }
    \label{fig:d3hex}
\end{figure}

This latest QECC encodes a logical qubit over a hexagonal lattice. As qubits are present on both the vertices and edges of the lattice,  the term heavy is used. This is a combination of degree-2 and degree-3 qubits hence there is a huge improvement in terms of average qubit degree in comparison with surface code structure which has qubits of degree-4 \cite{chamberland2020topological}. Fig. \ref{fig:d3hex} shows the lattice for a distance-3 heavy hexagonal code encoding one logical qubit. 
The heavy hexagonal code is a combination of surface code and subsystem code (Bacon Shor code) \cite{chamberland2020topological}.  A subsystem code is defined by $G$, a set of gauge operators where $\forall$ $g \in G$, $\ket{\psi} \equiv g\ket{\psi}$ \cite{bacon2006operator}. A gauge operator takes a codeword to an equivalent subsystem. In other words, a codespace in a subsystem code consists of multiple equivalent subsystems.  It is to be noted that the gauge operators are not necessarily commutative.  The product of two or more gauge operators forms a stabilizer, which keeps the codeword unchanged.

In  \cite{bhoumik2022efficient}, the authors propose a feed forward network based decoder to show that their decoder can decode a topological code, namely heavy hexagonal
code efficiently, in terms of threshold and pseudo-threshold, for different error models. Their machine learning based decoding method achieves ~$5\times $ higher values of threshold than that by
MWPM. The novelty of their work is exploiting the property of subsystem code to define gauge equivalence of errors which leads to reduction in the number of error classes. They have proposed two methods to  improve the threshold further by another 14\% by obtaining a quadratic reduction in the number of error classes for bit flip and phase flip errors.

Since training in machine learning can become expensive for higher distance codes, it may be possible to use a divide-and-conquer method to divide the error-correcting code lattice into multiple smaller (and possibly overlapping) sublattices so that each of them can be trained efficiently. However, this method has the challenge of \emph{knitting} the results from these sublattices into the final result corresponding to the lattice.

\section{Classical ML in Noisy Intermediate Scale Quantum (NISQ) era}
\label{nisq}
\begin{figure}[htb]
    \centering
    \includegraphics[scale=0.6]{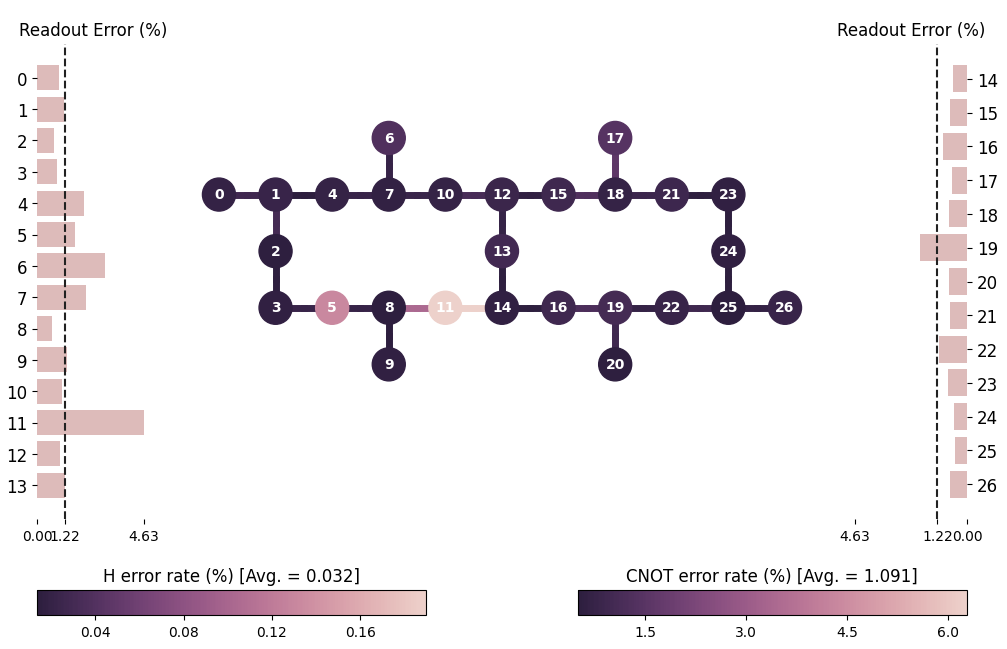}
    \caption{Noise profile of a 27-qubit IBM Quantum device}
    \label{fig:noise}
\end{figure}

Intermediate-scale quantum computers, having less than 1000 qubits, are available with many industry research labs such as IBM, Google, IONQ, etc. These devices are characterized by a small number of qubits, noisy gates and low coherent time., and  have been termed as Noisy Intermediate-Scale Quantum (NISQ) devices \cite{preskill2018quantum}. It is not possible to achieve fault-tolerance with these devices as (i) the number of qubits is not large enough, and (ii) the noise-profile of the device is still too high for concatenation to reduce the noise in the system \cite{majumdar2022fault}. In Fig.~\ref{fig:noise} we show the noise-profile of a 27-qubit IBM Quantum hardware. The threshold of surface code is $1\%$ \cite{fowler2012towards} and that of the heavy-hexagonal code \cite{chamberland2020topological}, which is designed specifically for the heavy hexagonal structure of IBM Quantum hardware, is $\sim 0.4\%$. As evident from Fig.~\ref{fig:noise}, the noise-profile of current quantum devices is much higher than the threshold of the quantum error correcting codes, and hence fault-tolerance using concatenation will not be useful as of now.

\begin{figure}[htb]
    \centering
    \includegraphics[height=6 cm, width= 9.5 cm]{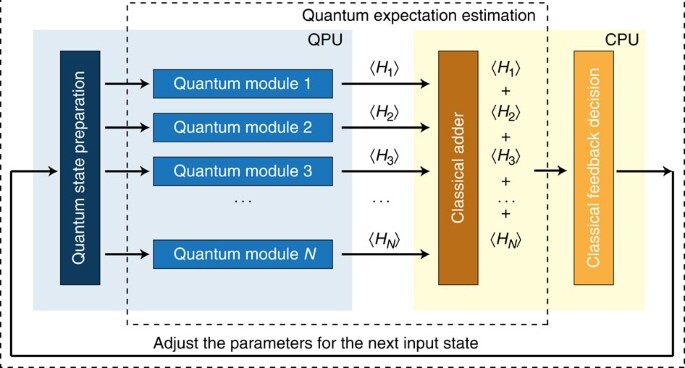}
    \caption{An overview of Quantum-Classical Hybrid Algorithms \cite{peruzzo2014variational}}
    \label{fig:nisq}
\end{figure}

While these quantum computers are not yet capable of general-purpose large-scale computing, researchers have studied hybrid quantum-classical algorithms which can be executed in these devices. These algorithms are, in general, divided into modules, some of which are outsourced to Classical Processing Units (CPU). Thus, the quantum circuit have shallow depth and is less susceptible to noise. In other words, these algorithms can still produce acceptable outcomes under noise. These algorithms find applications in Quantum Chemistry, Combinatorial Optimization, Quantum Machine Learning, etc. In Fig.~\ref{fig:nisq} we show a blueprint of the working principle of quantum-classical hybrid algorithms.

Interestingly, we find the application of classical ML models on these algorithms, as well as some of these algorithms themselves working as ML models. In the following subsections, we briefly touch upon some of these approaches.

\subsubsection{ML in Variational Quantum Eigensolver (VQE)}
Quantum-classical hybrid algorithms are shown to have applications in Quantum Chemistry, especially for finding the ground state energy of a system of molecules. These algorithms, which find an approximate solution to the ground state energy of molecular systems, are termed Variational Quantum Eigensolvers (VQE). The parameterized quantum circuit is termed as ansatz. The ideal requirement is that the ansatz should contain the solution to the problem. However,  it is not so mostly. The search space of the problem is usually exponentially large in the number of qubits. Therefore, an ideal ansatz that would contain the perfect solution would have a significant circuit depth and a large number of parameters. Increasing the number of parameters slows down the classical optimization, and hence the entire algorithm. Using a highly complicated quantum circuit makes it more susceptible to noise. Some standard ansatz circuits such as Unitary Coupled Cluster (UCC) are widely used for problems in Quantum Chemistry \cite{barkoutsos2018quantum}. This ansatz is designed from the knowledge of the problem domain, but does not consider the hardware connectivity of the qubits. A few later studies have looked into Hardware efficiency (HE) ansatz \cite{kandala2017hardware} for finding the ground state energy. These ansatzes ensure that the mapping of the circuit on the hardware does not result in too many SWAP gates.

The ideal scenario is to have a low-depth ansatz that provides a good approximate solution to the problem at hand.  In \cite{ostaszewski2021reinforcement}, the authors used reinforcement learning (RL) to incrementally obtain better ansatz having low depth but can still provide a good approximate solution. In Table~\ref{tab:rlvqe}, the authors showed that the RL-based ansatz has a lower gate cost, and depth compared to both HE and UCC ansatz. The authors took a trial over 10 steps, and the average is over all the trials. They showed that the energy estimate obtained is at par with the other ansatz, and in 2 out of 10 trials, they even obtained the perfect chemical accuracy.

\begin{table}[htb]
    \centering
    \caption{Comparison of gate cost and depth for different VQE ansatz}
    \begin{tabular}{|c|c|c|c|c|}
    \hline
        & Avg Depth & Min Depth & Avg \# Gates & Min \# Gates \\
        \hline
        RL & 14 & 12 & 36 & 29\\
        \hline
        HE & 17 & 17 & 63 & 63\\
        \hline
        UCC & 377 & 377 & 610 & 610\\
        \hline
    \end{tabular}
    \label{tab:rlvqe}
\end{table}

\subsubsection{ML in QAOA design}
VQE mostly finds applications in Quantum Chemistry. Quantum Approximate Optimization Algorithm (QAOA) is another family of hybrid quantum-classical algorithms that is aimed at finding good approximate solutions to combinatorial optimization problems \cite{farhi2014quantum}. QAOA is essentially a subclass of VQE where the ansatz design is governed by the Quadratic unconstrained binary optimization (QUBO) formulation of the combinatorial optimization problem. These algorithms ares characterized by a problem hamiltonian $H_P$ which encodes the problem to be solved, (e.g., Max-cut, minimum vertex cover), and a mixer hamiltonian $H_M$ which should anti-commute with $H_P$. A depth-$p$ QAOA is represented as:
\begin{equation}
    \label{qaoa}
    \ket{\gamma \beta} = \displaystyle \Pi_{l = 1}^{p} e^{-i \cdot \beta_l \cdot H_M}e^{-i \cdot \gamma_l \cdot H_P}\ket{\psi_0}
\end{equation}

where $\psi_0$ is the initial state (usually an equal superposition state), and $\gamma = \{\gamma_1, \gamma_2, \hdots, \gamma_p\}$ and $\beta = \{\beta_1, \beta_2, \hdots, \beta_p\}$ are the set of parameters. The objective of the algorithm is to maximize (or minimize) the expectation of $\braket{\gamma \beta|H_P|\gamma \beta}$. Upon obtaining the value of $\braket{\gamma \beta|H_P|\gamma \beta}$, the classical optimizer suggests a new set of parameters, and the algorithm is repeated with the new state $\ket{\gamma \beta}$. This process is repeated till convergence.

The time complexity of the algorithm is determined by the value of $p$. However, recent results show that the time duration of the classical optimizer is non-negligible \cite{guerreschi2019qaoa}. In fact, for larger problem size, it is the classical optimizer that takes up the majority of the absolute runtime of the algorithm. In order to avoid this issue, in \cite{alam2020accelerating} the authors proposed an ML-based technique for faster convergence of the classical optimizer. The ML aims to learn the correlation between QAOA parameters from smaller and larger values of $p$. After exhaustively finding the optimal parameters for lower depth, their ML model could predict a good starting point for the classical optimizer for higher $p$ so that the optimizer could converge quickly. The authors used different ML methods, of which Gaussian Process Regression (GPR) provided the best result. With this ML model, they could reduce the runtime of the classical optimizer by $44.9\%$ on average. Their result is over 330 different graphs and 6 QAOA instances for each graph.

This finding is later supported by \cite{akshay2021parameter} who showed that the optimal parameters of a QAOA instance concentrate. In other words, if the optimal parameters for problem instances with $n$ and $n+1$ qubits are $\{\gamma_n, \beta_n\}$ and $\{\gamma_{n+1}, \beta_{n+1}\}$ respectively, then, :
            \begin{center}
                $\exists l > 0$ $|\beta_{p+1} - \beta_{p}|^2 + |\gamma_{p+1} - \gamma_p|^2 = \mathcal{O}(\frac{1}{n^l})$
            \end{center}

This paper depicts that the ML model in \cite{alam2020accelerating} is indeed learning this concentration of parameters.
The original QAOA proposed by Farhi et al \cite{farhi2014quantum} was for unconstrained optimization problem. Later on, Hadfield proposed a variant of QAOA for constrained optimization problems as well \cite{hadfield2019quantum}. This approach is similar to the original QAOA, except for the mixer Hamiltonian, which now becomes more complicated to ensure that a valid solution is mapped into superposition of valid solutions only. Modification of the initial state \cite{egger2021warm}, mixer hamiltonian \cite{zhu2022adaptive} and problem hamiltonian \cite{majumdar2021optimizing} have been proposed to obtain faster convergence or reduce the effect of noise on QAOA. However, the ML method to predict the parameters holds good for these variants as well.

\subsubsection{Variational Approach to Error Correction}
Previously, we have briefly touched upon classical ML methods to further optimize the VQE and QAOA algorithm. However, now we show that these ansatz have the ability to act as ML models themselves. Near-term quantum devices cannot afford quantum error correction due to requirement of a large number of qubits to achieve error correction and fault tolerance. Therefore, error mitigation techniques have been proposed in the literature. These methods cannot nullify, but can minimize the effect of errors.

\begin{figure}
    \centering
    \includegraphics[height= 8 cm, width= 15 cm]{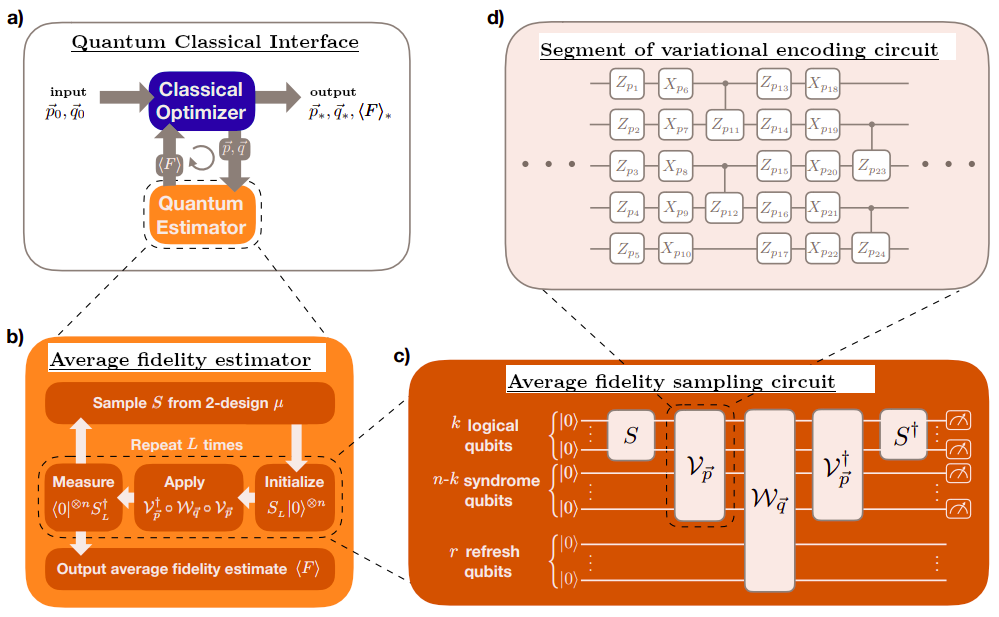}
    \caption{A schematic diagram of the variational QECC proposed in \cite{johnson2017qvector}}
    \label{fig:qvector}
\end{figure}

\begin{figure}
    \centering
    \includegraphics[height=6.2 cm, width= 8 cm]{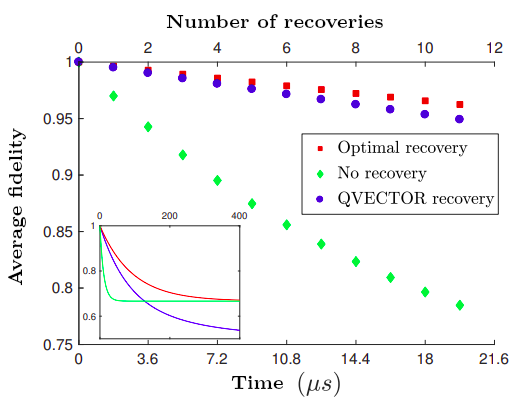}
    \caption{Average Fidelity as a function of time for the variational QECC, optimal QECC and no error correction}
    \label{fig:compqvector}
\end{figure}

There are different sources of errors, and error mitigation techniques are usually aimed to reduce the effect of a few of them. In \cite{johnson2017qvector}, the authors proposed a method using hybrid quantum-classical algorithms to reduce the overall effect of the error on the quantum circuit. Their motivation was not to reduce each error but rather to learn the overall effect of error on the system, and then minimize it. Therefore, this leads to a model-free method of reducing the effect of noise.
In Fig.~\ref{fig:qvector}, we show the schematic diagram of the variational QECC proposed in \cite{johnson2017qvector}. The system is initialized in a 2-design circuit, which is a random set of Clifford gates\cite{}. This is followed by an encoding by a parameterized circuit $V_{\Vec{p}}$. The system undergoes noisy evolution by a $W_{\Vec{q}}$, and the fidelity $\braket{0^{\otimes n}| S^{\dagger} V_{\Vec{p}}^{\dagger} W_{\Vec{q}} V_{\Vec{p}} S |0^{\otimes n}}$. The classical optimizer tunes the parameter to maximize the Fidelity.  Fig.~\ref{fig:compqvector} depicts how the authors establish that this proposed variational model provides a performance which is almost the same as the optimal recovery technique, e.g., the 5-qubit QECC \cite{laflamme1996perfect}. Therefore, this model of variational algorithm readily demonstrates that it can learn the underlying noise model of the channel (which is unknown) and tune the parameters to propose a recovery scheme that is comparable to the optimal QECC.

Machine learning can play a big role in QECC by predicting the best QECC for use under a current noise scenario. An approach for this has been explored in \cite{su2023discovery} for biased Pauli noise. It will be interesting to see whether it is possible to further fine-tune this method by learning the noise in the system \cite{van2023probabilistic}.

\section{Quantum Machine Learning}
\label{qml}
Both machine learning and quantum computing have progressed hand-in-hand, each benefiting from the other. In this section, we briefly touch upon the usage of quantum computation to search for more powerful and efficient machine learning models. Quantum Machine Learning (QML) is a contemporary theoretical field that resides at the intersection of quantum computing (QC) and machine learning (ML).

In classical machine learning, to model any problem, our target is to find a function $f$, given $x$ and $y$, such that: $y = f(x)$. We provide the input and expected output (label) to a ML model, which learns the rules without explicitly telling the computer how to solve the problem. The program learns to do so itself. A loss function (mathematical expression which calculates the amount by which the algorithm has missed the accurate target) checks on how correct a machine learning solution is. During ML model training, it is obvious that at the beginning not all of them are correct. QML also targets to minimize the loss function using a property called Quantum Tunneling (QT). QT searches through the loss function space fully to find the value with minimum loss. There are some popular methods that QC uses to solve ML problems, such as Quantum Neural Network (QNN), Quantum Principal Component Analysis (QPCA),
Quantum Support Vector machines (QSVM), Quantum Reinforcement Learning (QRL),
Quantum Optimization (QO) etc. 

Ideally, a QML algorithm may work on classical or quantum data. When a QML algorithm works on classical data, it becomes necessary to encode the data into qubits, which are fed to the QML. Therefore, we first discuss a few widely used encoding methods that are utilized to encode the information of classical data into qubits. Then we follow up with some algorithms such as QNN, QSVM and talk about two practical usage of QML in classifying facial expression \cite{mengoni2021facial} and handwriting from MNIST dataset \cite{jiang2021machine}.

\subsection{Encoding classical data into qubits}
A quantum system cannot directly read from classical data. So it is necessary to encode the information of the data into qubits. Note that this idea of \emph{encoding} is not to be confused with the \emph{encoding} in quantum error correction. Here, encoding simply implies uploading the classical data on the qubits. While there is no steadfast rule on the nature of encoding, the following three methods are used most widely.

\begin{enumerate}
    \item \textbf{Basis Encoding}: The classical data is represented as a sequence of bits (similar to One Hot Encoding in classical machine learning). For example, if $x \in \{0,1\}^n$   is the One Hot Encoding of a particular classical data, its corresponding quantum encoding is $\ket{x}$. For two distinct $\ket{x} \neq \ket{x'}$, we have $\braket{x|x'} = 0$.
   Let $x_1, \hdots, x_m$ be the $m$ classical training inputs. Then the corresponding quantum data is of the form
    $$\ket{\psi}_{init} = \frac{1}{\sqrt{m}}\displaystyle \sum_{i=1}^m \ket{x_i}.$$
    This method is perhaps the simplest form of encoding classical data in a quantum device. Nevertheless, it often requires a significant number of qubits, especially if fractions are to be encoded, and thus defeats the purpose of using fewer qubits.
    
    \item \textbf{Amplitude Encoding}: The value of the classical data is encoded as the probability amplitude of the basis states. For example, if a classical data is represented as $\{\alpha_1, \alpha_2, \hdots, \alpha_k\}$, $\alpha_i \in \mathbb{R}$ $\forall$ $i$, then the corresponding quantum input is
    \begin{center}
    $\ket{\psi} = \displaystyle \sum_{i=1}^k \frac{\alpha_i}{|\alpha_i|^2}\ket{x_i}$
    where $x_i \in \{0,1\}^n$.
    \end{center}
    \item \textbf{Angle Encoding}: The values of the features   can be encoded into the angles of the amplitude of qubits as well. Consider a dataset with $2n$ features. This method requires only $n$ qubits to encode $2n$ features, and the qubits can be implemented with low depth circuits. The values $x_i$ and $x_{i+1}$ for features $i$ and $i+1$ respectively can be encoded into a single qubit as
    \begin{center}
        $\ket{\psi}_{i,i+1} = cos(x_i)\ket{0} + e^{i \cdot x_{i+1}}sin(x_i)\ket{1}.$
    \end{center}

\end{enumerate}

Till now, no encoding system provably outperforms the others for QML. Researchers have used one or more of these encoding methods according to the problem at hand. After encoding the data into qubits, the \emph{QML algorithm} tries to learn from that data to generate the desired results. Next, we  discuss two broad classes of QML algorithms, namely Quantum Neural Network, and Quantum Support Vector machine.

\subsection{Quantum Neural Network (QNN)}\label{qnn}
QNN is a class of ML models that are used in quantum computers. These deploy quantum properties of entanglement, superposition, and interference for the computation. 
In \cite{abbas2021power} the authors study the trade-offs, i.e., whether QNNs are more "powerful" than classical NNs. They found that   well-designed QNNs are able to achieve a higher capacity captured by the effective dimension,  and faster training ability influencing the Fisher information-theoretic property  than comparable classical feed forward NNs. But, how do we define this term "powerful"? 

In a classical NN, we can naively count the number of parameters of our model. A higher number of parameters can capture more information about the relationship between the data and variables of the model. On the other hand, all the parameters may not be useful. Another popular method to estimate the power is the VC (Vapnik–Chervonenkis) dimension \cite{vapnik2013nature}. It measures the model capacity, expressibility and complexity, and derives the error bounds on how well a model generalises (i.e., performs on unseen data). But although it has attractive properties in theory,  in practice determining the VC dimension is difficult, as the generalized bound is loose in case of deep NNs. There is a third metric, named as the Effective-dimension, which captures the size of the ML model in a higher dimensional space, rather than simply counting the number of parameters. It estimates the size which a model holds in model space where the Fisher information matrix  \cite{rissanen1996fisher} serves as the metric (instead of concentrating on parameter space like classical ML). This shift from the parameter space to the model space, is dependent on the Fisher information $F(\theta)$, which gives us a notion of distance in the model space, and   $\sqrt{\det{F(\theta)}}$ is the actual volume in the model space.  Fisher information is a way of quantifying the amount of information that an observable random variable $R$ carries about an unknown parameter $\theta$ of a distribution that models $R$. In other words, it is the variance of the  expected value of the observed information.  Hence the effective dimension is used to look into the size of the model space. Moreover this effective dimension is a data dependent notion; hence more data  means that the model space can be observed more clearly. 

In classical ML, the error of the model on new data is called the generalization error. A capacity measure gives a bound on this generalization error. In \cite{abbas2021power}, the authors proved that this generalization error can be bounded by the effective dimension as
$P(\displaystyle \sup_{\theta \in \Theta}|error_{true} - error(n)_{emprirical}| \geq \epsilon) \leq Capacity(|\Theta|,n)$,
where $error_{true}$ is the inherent true error, $error(n)_{emprirical}$ is the empirical error which we approximate using the data. The difference between these two is the generalisation gap, called generalization error, and is bounded by the effective dimension which is $Capacity(|\Theta|,n)$. It quantifies on the size of the model space $|\Theta|$ and on  a datapoint $n$. It was also shown that with increasing noise (randomness) in the data, the effective dimension increases. Hence it is  able to accurately capture the generalisation behaviour of an ML model.

\begin{figure}[hbt]
    \centering
    \includegraphics[height=5 cm, width= 15 cm] {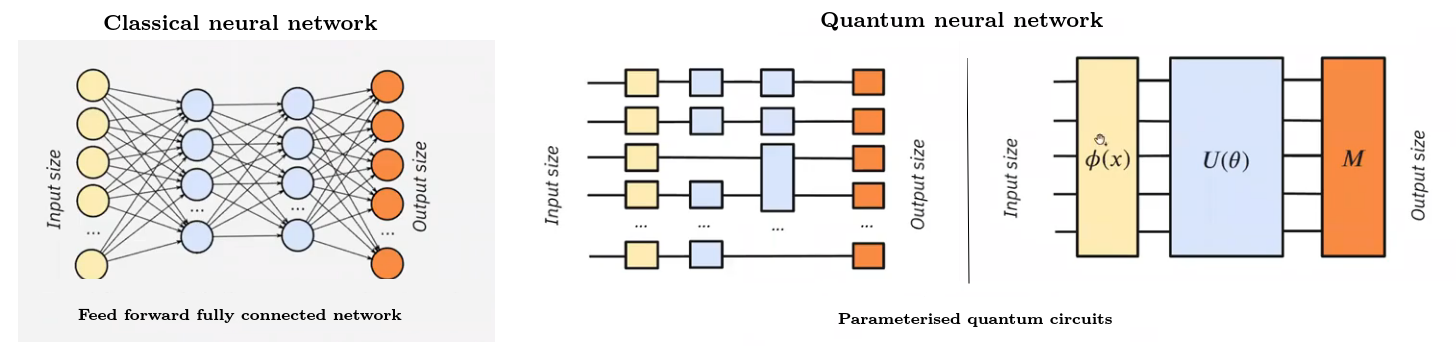}
    \caption{Classical Neural Network vs Quantum Neural Network }
    \label{fig:ann_vs_qnn}
\end{figure}

QNNs are a subclass of variational quantum algorithms, which consists of quantum circuits that contain parameterised gate operations. We can see a schematic diagram comparing the classical and quantum NN in Figure \ref{fig:ann_vs_qnn}, and these can be comparable if these have same  size of input and output, and the number of  trainable parameters.

\begin{figure}[hbt]
    \centering
    \includegraphics[height=5 cm, width= 13.5 cm] {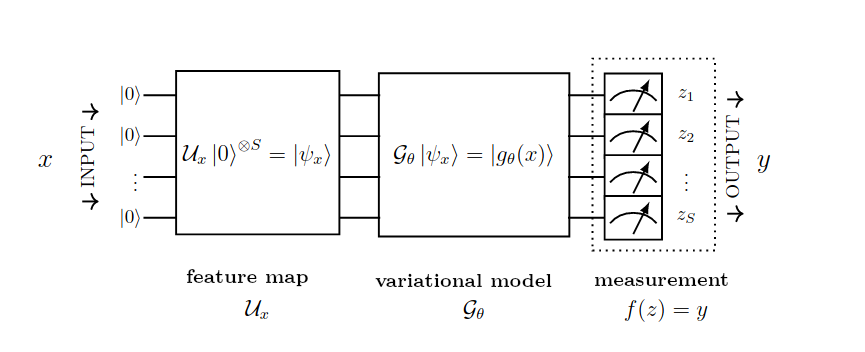}
    \caption{Quantum Neural Network: an overview \cite{abbas2021power}}
    \label{fig:qnn}
\end{figure}

In a QNN, $\phi(x)$ is the feature map that encodes the data \cite{havlivcek2019supervised}. Then comes the variational part of the circuit which consists of $CNOT$  and $R_Y$ gates. Next, there is post processing where every qubit is measured in the $\sigma_Z$ basis and the parity of the output bit strings are checked to map the corresponding probabilities to certain labels. In \cite{abbas2021power}, the authors have shown that between a classical model and a quantum model of neural network with same sizes of input and  output and parameter space, the effective dimension of the QNN is much higher than its classical counterpart.

A very simple and  easy quantum model  has a straight forward feature map and data encoding strategy. If there are four features in the feature vector $(x_1,x_2,x_3,x_4)$, these are encoded with Hadamard gate followed by an $R_Z$ rotations with angles $(x_1,x_2,x_3,x_4)$ respectively. But in a proper QNN after mapping the feature values using Hadamard and $R_Z$ rotations, there exist entanglement between the qubits, and higher orders of the data (product of the feature values) are also encoded. Hence they proved that the effective dimension of this QNN is much higher than the easy quantum model.

In ML, the Hessian matrix is the second derivative of the cost function which is to be minimized, and provides the intuition of the curvature of the landscape of the ML model to calculate the minima. This Hessian matrix coincides with the Fisher information matrix under certain conditions. An inherently quantum phenomenon, named as barren plateau, is a trainability problem that occurs in machine learning based optimization algorithms when the search space turns flat.  Hence the algorithm cannot find the downward slope in the landscape and there is no clear path to the minimum  due to the gradients vanishing. As a result, the entries to the Hessian matrix also goes to zero. This idea can be extended to Fisher information matrix also! They have shown that a model suffering from a barren plateau has a Fisher information spectrum with an increasing concentration of eigenvalues approaching zero as the number of qubits in the model increases, which happens in the classical NN. Conversely, a model with a Fisher information spectrum that is not concentrated around zero is unlikely to experience a barren plateau, which is the case for a QNN. Lastly, via numerical simulations they have shown that the loss for a QNN is much less (39\%) than the classical counterpart. 

For efficient  Artificial Neural Network (ANN) architectures, realization of
newly designed hardware devices which exploit the inherent parallelism in the NNs are necessary. These are known as Hardware  Neural  Networks  (HNN) \cite{misra2010artificial}. In case of classical NNs,  we have observed  a significant development directly
in terms of hardware over a long period. One of the most widely used geometry of hardware architecture is known as  reservoir network. Here the weights of all the connections are controlled by a large set of randomly connected nodes called reservoir. The input is fed to the reservoir with random connections, and only the weights of the output layer are controlled. For creating QNNs, the  classical reservoir network nodes are replaced with quantum nodes, hence the reservoir becomes a quantum physical system and the input and output data can be either classical or quantum information. Here they use a RNN to make the feature space, or reservoir. The connection between the reservoir output and the
final output is used for training (minimization of cost function). The training, therefore happens entirely  outside of the reservoir, making it easy for physical implementation. The quantum systems have a huge Hilbert space, which can provide the quantum reservoir a high memory capacity. Currently quantum reservoir computers are used to couple quantum states either in the input or at the output. It is still an open problem if both input and output can be in quantum states. \\

\begin{figure}[hbt]
    \centering
    \includegraphics[height=5 cm, width= 8 cm] {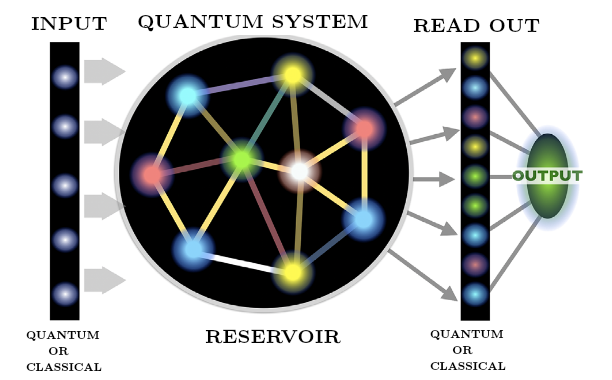}
    \caption{Quantum Reservoir Network \cite{misra2010artificial}}
    \label{fig:QRN}
\end{figure}

Using this reservoir technique, in \cite{krisnanda2021creating} the authors have shown that the QNNs can be used for quantum information processing (QIP) tasks (eg., quantum cryptography, quantum secret sharing, quantum memory, etc.) in a novel way. QIP needs exotic quantum states as a basic necessity, which are usually created with versatile methods customized to the determined set of resource states. The authors proposed an adaptable integrated scheme for the preparation of state which is based on a driven quantum network and is made of randomly-coupled fermionic nodes. The outcome of this kind of system is overlaid by a linear mixing where the model trains phases and weights to get the output quantum states. They have shown that their method is robust, and this method can create near-perfect maximally-entangled states and other states like GHZ states \cite{greenberger1989going}, W \cite{dur2000three} and NOON \cite{kok2002single} states. The authors have also considered the noisy environments such as energy decay, dephasing, and depolarization to get the target state with high fidelity ( $\sim$ 0.999), up to a permissible limit. Beyond that limit, a method is proposed for concentrating entanglement by mixing with other states which are present in a bigger network. This system has the same quantum network explained above as a quantum reservoir, which is created by few
fermionic nodes (eg., quantum dots) which interact with each other with random coupling strengths.

\subsection{Quantum Support Vector machine}
\label{qsvm}

As mentioned earlier, for classification using a support vector machine, the kernel is an important concept because data cannot simply be separated by a hyperplane in the original space of the data points. Hence, there is the need of non-linear transformation functions.
There are few classification problems that need a feature map for which the kernel cannot be computed efficiently in  a classical setup as it needs large computational resources (exponential with the size of the problem).

In \cite{havlivcek2019supervised}, the authors have shown that this issue can be resolved on a quantum processor by estimating the kernel in the feature space. For classification by a quantum computer, first the classical data point $X$ are mapped to a  quantum datapoint $\ket{\phi(X)}$ by the circuit $V \phi(X)$, where $\phi(X)$ could be any classical function applied on the classical data $X$. Then, a parameterized quantum circuit $W(\theta)$ with parameter $\theta$ that processes the data, is needed. Finally, a measurement circuit is applied that returns a classical binary value for each classical input $X$ to identify the class label. They propose two SVM-type classifiers to process the provided data classically. Then for obtaining the quantum advantage, the  quantum state space is used as the feature space, by non-linear data mapping to a quantum state. 

One approach uses a variational circuit for generating the separating hyperplane. Another approach uses a quantum processor to calculate the kernel function to implement a conventional Support Vector machine. They use self-generated artificial data so that it can be properly classified by the feature map, consisting  of 20 data points per class for both training and testing. In the result, they have shown that even for noisy datasets it can achieve a success rate of 100\%. Although the main question in these quantum-enhanced feature spaces is the amount of \emph{enhancement}. Only because a task is hard in classical computers does not make the quantum version advantageous. The open problem is to find cases to prove the advantages of this (or other QML) protocol. So if a quantum feature map is chosen that is hard to simulate with a classical computer, then quantum advantage might be obtained.

In \cite{zhang2023quantum}, the authors proposed a quantum support vector machine based on amplitude estimation (AE-QSVM) which does not have the constraint of repetitive process and thus saves the quantum resources. Although there is extensive research work going on in the domain of quantum machine learning, there are many questions regarding the prospective of machine learning on quantum computers that are still unanswered. For example, does the current QML algorithm work better than classical in practice?

In \cite{simoes2023experimental}, the authors survey how QML can be used for solving small hands-on problems. They present the experimental analysis of kernel-based quantum SVM and QNN using 5 different datasets. They show that quantum SVM outperforms their classical counterparts on average by ~4\% in accuracy both on a  simulator and real quantum machine. QNN executed on a quantum computer outperforms quantum SVM on average by upto ~5\% and classical neural networks by 7\%.

In \cite{mensa2023quantum}, the authors propose a general-purpose framework combining a classical Support Vector Classifier algorithm with quantum kernel estimation for ligand-based virtual screening (LB-VS) on real-world databases, for discovering new drugs in a faster and cost-effective manner, especially for emerging diseases such as
COVID-19. In.  They show that it performs 13\% better than the classical counterpart.

\subsection{Quantum Reinforcement Learning}
\label{qrl}
Classical reinforcement learning models are sensitive to errors during training  \cite{wang2020reinforcement}, hence a robust reinforcement learning framework that
enables agents to learn in noisy environments is needed.
The authors in \cite{dong2005quantum} presented the idea of Quantum Reinforcement Learning (QRL) inspired by the state superposition principle and quantum parallelism. In \cite{dong2008quantum}, the authors give a formal QRL algorithm framework. They demonstrate the advantages of QRL for faster learning and obtaining a good tradeoff between exploration and exploitation through simulated experiments.

Reinforcement learning with VQAs has been proposed in \cite{chen2020variational} for an error-free environment. The results are similar to neural networks on small classical benchmark tasks \cite{skolik2022quantum}. In \cite{skolik2023robustness}, they address  the effect of training quantum reinforcement learning models under the influence of hardware-induced noise, and report the effect on the performance of the agents and on the robustness of the learned policies.

\subsection{Quantum Image Processing}
  Nowadays classical image processing is widely used in commercial sectors such as autonomous vehicles \cite{li2004springrobot}, facial recognition \cite{guo2001support}, motion detection and object recognition \cite{pal2020granulated, carion2020end} etc. Quantum image processing (QIP) employs the advantage of quantum mechanical properties for representing pixels of the image in a quantum computer. Depending on the image format in the quantum computer, various image operations can be implemented. QIP can have certain advantages over classical image processing, because it can exploit the quantum parallelization which inherently comes from superposition and entanglement.  It holds the promise of substantial speed-up for a few common operations like edge detection \cite{zhang2015qsobel}.
  
  Facial Expression identification is an important job needed for human-computer interaction in different scenarios. It comprises a classification problem that categorizes face images with different expressions such as happy, sad, angry, scared, etc. The dataset of the human face is extremely heterogeneous due to facial features, different poses, and also the background. Classical Facial Expression identification consists of image pre-processing followed by feature extraction and classification of the expression. In \cite{mengoni2021facial}, the authors have described the steps for the same on a quantum computer. The first step is image pre-processing, which is done classically using the  FFHQ dataset \cite{styleganversions}. During feature extraction, the pre-processed images are mapped into graphs. The classification step is mainly the quantum part where the features are mapped into the amplitudes of the quantum state, which in turn forms the input to the quantum circuit (by using a technique like the nearest centroid method) for representing the facial expression image classifier, depending on the Euclidean distance on the graphs. The authors compare the results with the classical algorithm. When there are four vertices, the classical method achieves a 99\% accuracy, while the quantum counterpart achieves an 88\% accuracy for a complete graph. This results in a gap of 11\%. However, as the number of vertices increases to 20, both the classical and quantum methods achieve a 100\% accuracy for complete graphs. Consequently, this gap gradually diminishes to 0\%.

 They have also mentioned that if the graph dimension is of higher magnitude than a complete graph, classification may not be feasible. Instead, a meshed graph strategy can be used with a trade-off in terms of accuracy. 
  
\subsection{Quantum accelerator for ML
}
\label{aspdac}


There has been substantial progress in studies on the acceleration of neural networks on classical processors, e.g. CPU, GPU, ASIC, FPGA over decades. But with the increasing scale of the application, a bottleneck with respect to memory comes in place which is known as memory wall. This is where advanced quantum computing comes as a solution. The development of machine learning using a classical hardware accelerator can be done in two phases:
\begin{itemize}
    \item Neural network tailored hardware design:  FPGA-based DNN accelerators \cite{jiang2019achieving}, FPGA-based cost-optimal design for timing-constrained CNNs \cite{jiang2018heterogeneous}, etc.
    \item Neural network and hardware accelerator co-design: in \cite{bian2020nass}, the authors proposed an integrated structure that can search for tailored neural network architectures designed precisely for secure inference (SI). In \cite{cai2018proxylessnas}, the authors proposed a "Proxyless - NAS (Neural architecture search)" which is capable of learning the architectures for large-scale target tasks and target hardware platform directly(without any proxy).
\end{itemize}   

 In \cite{jiang2021machine}, the authors stated that the neural network co-design and quantum circuits design
 must be executed
to fully utilize the potential of the quantum computer (QC).
The full acceleration system consists of three units:
\begin{itemize}
    \item Pre- and post-processing of the data on a classical computer
    \item NN accelerator on the quantum circuit including quantum state preparation ($U_P$) 
    \item QC-based neural computation ($U_N$).
\end{itemize}  
\begin{figure}[hbt]
    \centering
    \includegraphics[height=5 cm, width= 8 cm] {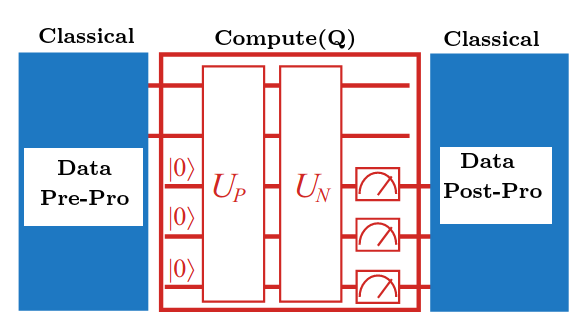}
    \caption{Quantum Computing Accelerator \cite{jiang2021machine}}
    \label{fig:ML_QC}
\end{figure}

Preparation of the quantum data to quantum state encoding is the initial step. If the first column of a unitary matrix $U$ encodes the vector $U_0$ of $2^N$ data, then $U_0 = U\ket{\psi}$ where $\ket{\psi} = \ket{0} ^{\otimes N}$ is the initial state. The next step of quantum-state preparation has the potential of affecting the complexity of the whole circuit significantly. The authors propose the use of a quantum memory (qRAM) where a binary-tree-like structure stores the vector in $U_0$. This is used to query in quantum superposition and to generate the states in a competent way. Next, they apply the popular MNIST data set \cite{mnist} with the goal of carrying out a case study where the image data (16 inputs) is encoded onto 4 qubits. In the neural computation part, which is the primary element in the implementation of QML, the weighted sum with quadratic function is calculated using the binary weights $W$. 

The computation of the hidden layer consists of two parts. The first is to multiply inputs and weights (quantum gates such as $X$ gate and  3-controlled-$Z$ gate accompanied by three trigger qubits are used to operate the weights with the inputs). The second is to apply the quadratic function on the weighted sum (the Hadamard ($H$) gates are applied on each of the qubits to accumulate all states to the state-$0$ followed by swapping the amplitudes of state-$0$ and state-$1$, and then applying the N-control-$X$ gate for the extraction of the amplitude to a single output qubit $O$, in which the probability of $O=\ket{1}$ which is the square of the weighted sum). Then they finally get the output layer for the final results. With these $N$ output qubits, these qubits are used continuously to directly compute the outputs. However, there is a need to modify the fundamental computation to the multiplication of random variables, as the probability of the qubit to be in state $\ket{0}$ is associated with the data represented by a qubit. Running a simulation or execution on the IBM Quantum processors can measure the output qubits, and the classification  results can finally be obtained.

\section{Classical machine learning in quantum communications and cryptography}
\label{qcomm}

\begin{figure}
     \centering
     \begin{subfigure}[b]{\textwidth}
     \centering
    \includegraphics[height=4.1 cm, width= 13 cm] {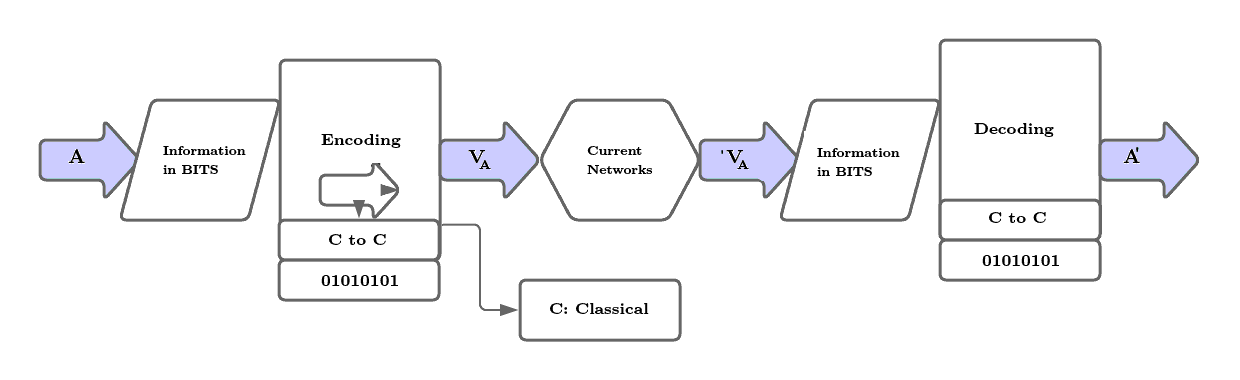}
   \caption{}
    \label{fig:cvsqc}
     \end{subfigure}
     
     \begin{subfigure}[b]{\textwidth}
     \centering
  
    \includegraphics[height=5.1 cm, width= 13 cm] {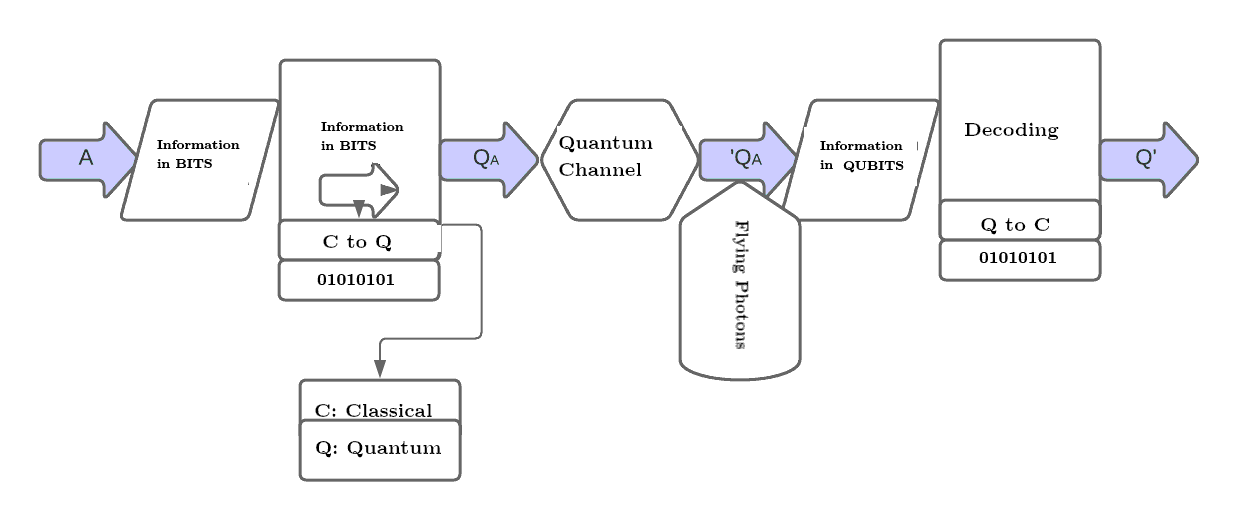}
   \caption{}
    \label{fig:cvsqc1}
    \vspace{-0.3cm}
     \end{subfigure}

    \caption{(a) Classical and (b) Quantum Communication Systems}
    \label{fig:qccc}
    
\end{figure}

Transmission of classical or quantum information through a quantum communication system from one location to another using a quantum channel is known as a quantum communication system. The information is usually transmitted as a photon along optical cables between sender and receiver, and is capable to move at a speed close to that of light, along with weak environment interaction. Photon communication is prone to challenges with optical light sources. Additionally, even in nonlinear materials, two photons do not interact strongly. Fig. \ref{fig:qccc} narrates the communication of classical bits and quantum qubits. Its main applications include cloud computing, cryptography-related tasks, secured storage, and other secure communications \cite{Buchmann2017parameter}, as well as a secure and safe quantum network for distributing quantum properties such as randomness, entanglement and non-locality at connected but remote locations. Quantum communication \cite{Zoller2005parameter} protects data by employing quantum mechanics  and allowing particles to be superimposed. 

There are various techniques for implementing quantum communication. 
Some of the current approaches being researched include universal quantum gate model, analog quantum model and quantum annealing.
Several sectors are progressing in quantum communications, for example, the first commercial quantum annealing device was developed by D-wave \cite{Dwaave2021parameter}.
Technologies for quantum communication systems rely on the development of diverse central protocols and schemes, such as quantum cryptography \cite{zhao2014apply, Gottesman2003apply}, teleportation \cite{Bennett1993apply} and to handle fundamental problems like channel noise accumulation and decoherence such as quantum repeaters \cite{Bennett1996parameter} and entanglement purification to enable for scalable long-distance quantum communication. In \cite{Julius2020apply} the authors have shown how ML is used to address the prime areas of teleportation, purification of entanglement, quantum repeaters, and quantum protocols. 



Low data rates, quantum channel and its security are the integral issues of quantum communication till date and hence  classical machine learning can be applied to discover the properties of free space quantum channels. The authors of \cite{Ismail2019apply} have proposed a supervised ML technique to discover the atmospheric strength of free space quantum channel in the form of Strehl ratio. The study has revealed that the application of the random forest method has resulted in a good prediction of the Strehl ratio of a quantum channel with less mean error.
 
\subsection{In quantum cryptographic protocols}
Quantum Key Distribution (QKD) is the new secure mechanism of communication to share keys secretly among communicating parties. 
The process of QKD involves a strong secure key exchange system to encrypt and decrypt information. The classical key distribution techniques depend on public key ciphers by using convoluted mathematical computations and hence demand more processing power to decipher. These ciphers also face numerous challenges such as weak random number generators and continuous new strategies for attack. Unlike classical mathematically oriented key distribution, QKD involves  basic properties of quantum mechanics to protect data. 
According to the no-cloning theorem \cite{nielsen2002quantum}, no  identical copy of an arbitrary unknown quantum state can be created and hence it is very difficult to identify or to copy data across two ends. In a sifting phase, Alice and Bob post their slots of detection, and both Alice and Bob are set to detect the polarization of schemes. At any point of time when Bob measures the corresponding measurement of Alice, then it is in principal state, else whenever any attacker peeps or attempts to disturb the system, the system  changes automatically so that the intruder ends up with failure to decrypt. 

The transmitted photon reaches the destination through a beam splitter that guides the photon to take any random path to reach the photo collector. The receiver confirms the photon sequence to the sender after receiving the photons and these are verified across the sender and receiver. If any photon is received from the wrong beam splitter, then those are discarded otherwise the secret key is exchanged using the bit sequence. The final secret key is exchanged safely by adopting a technique that delays the privacy of amplification and this post-processing method erases all the information from any eavesdropper who has acquired any information about the key.

\begin{figure}[hbt]
    \centering
    \includegraphics[height=8 cm, width= 15 cm] {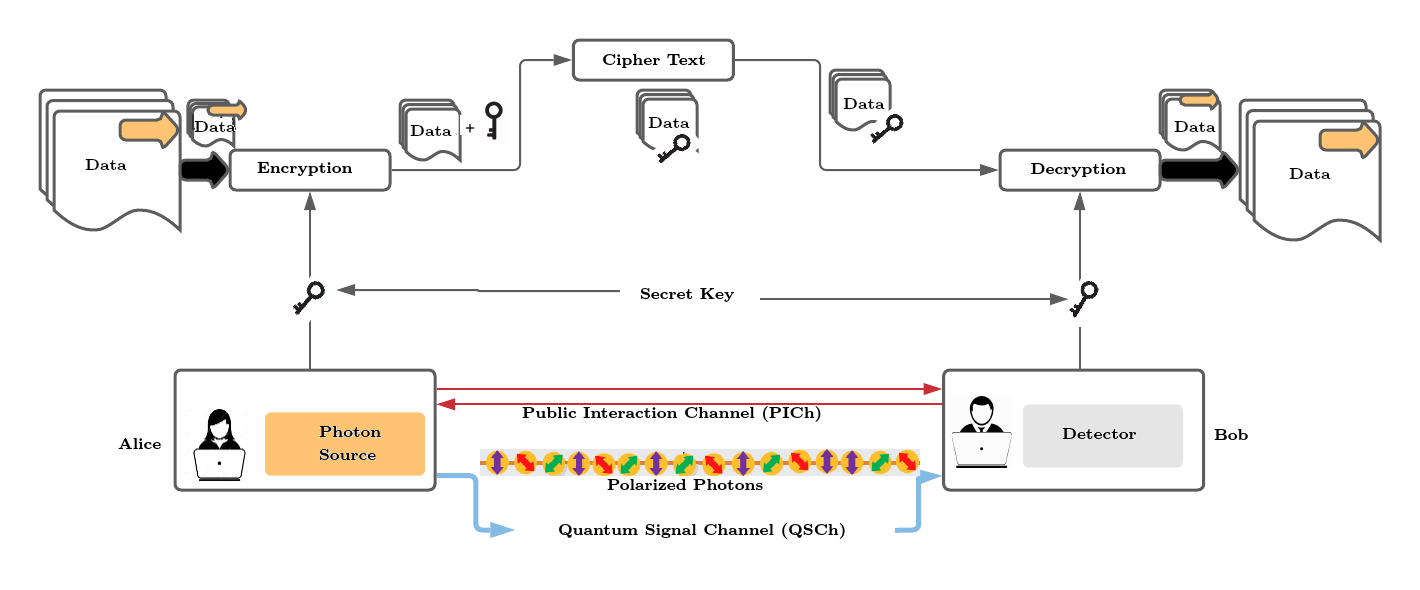}
    \caption{Basic QKD System {\tiny (Image source: \cite{Purvasharma2019apply} )} }
    \label{fig:basicqkd}
\end{figure}

Figure \ref{fig:basicqkd} shows the basic QKD system and the mechanism of exchanging secure information. In general a QKD system is equipped with two types of channels, namely PICh (public interaction channel) and QSCh (quantum signal channel) along with encryption, decryption blocks, and a QKD protocol. At the outset, QKD protocol is used to initiate a secure interconnection between Alice and Bob, also used to generate the secret keys and to analyze the accurate information split between sender and receiver while generating the keys. 

\begin{table}[htb]
    \centering
    \caption{QKD Protocols:  Attacks  and Salient Features }
    \begin{tabular}{|p{2cm}|p{2.5cm}|p{1cm}|p{0.8cm}|p{0.8cm}|p{5cm}|}
    \hline
 \multicolumn{3}{|c|}{ \textbf Protocol}   & \multicolumn{2}{|c|} {\textbf Attacks} & 
\multicolumn{1}{|c|} {\textbf Salient Feature}\\ \hline
  \textbf  Family  & \textbf Name & \textbf Scheme  & \textbf {PNS} & \textbf {DoS} & \\
     \hline
   \multirow{5}{*} {DV-QKD \cite{Ivan2020apply} } & BB84 \cite{Bennett1984apply}, 1984 & P-M  & V & V & First QC protocol with 4 polarization states\\
  \cline{2-6}
   & E91 \cite{leilei2018apply}, 1991 & E & V & V & First Entanglement protocol with 2 non-orthogonal states \\
  \cline{2-6}
   & B92 \cite{mendonca2007apply}, 1992 & P-M  & V & V & Similar to BB84, uses 2 non-orthogonal states \\
  \cline{2-6}
     & SSP \cite{Bechmann1999apply}, 1998 & E  & V & V & Uses 6 polarization states \\
  \cline{2-6}
     & SARG04 \cite{scarani2004quantum}, 2004 & P-M  & V & R & Difference in classical phase of BB84 \\
  \hline
   \multirow{2}{*} {CV-QKD \cite{Ivan2020apply} } & Discrete modulation protocol \cite{Shouvik2019apply} – BB84 2000 & P-M  & R & R & Updated BB84, includes  squeezed-state and discrete modulation\\
  \cline{2-6}
   & Gaussian protocol \cite{Christian2012apply} – BB84 2001 & P-M  & R & R & Updated  discrete modulation protocol with  Gaussian modulation\\
  \hline
  \multirow{2}{*} { DPR-QKD \cite{Mhlambululi2020apply} } & DPS \cite{Usuga2016apply}, 2003 & P-M  & R & R & First DPR based QKD protocol, uses one bit delay circuit to create and measure qubits\\
  \cline{2-6}
     & COW \cite{Damien2007apply}, 2004 & P-M  & R & V & Similar to DPS but uses pulses for creating photon, and bit encoding is done using a sequence of one non-empty (µ)-pulses.\\
  \hline
  
  \hline
   \multicolumn{6}{|c|}{
     SSP: Six State Protocol, P and M: Prepare and Measure, E: Entanglement based}\\
    
    \multicolumn{6}{|c|}{V : Vulnerable    R : Robust, PNS: Photon Number Splitting, 
    DoS: Denial of Service}\\
   

  \hline
  \end{tabular}
  
    \label{tab:dvqkd}
\end{table}

\begin{figure}[hbt]
    \centering
    \includegraphics[height=8 cm, width= 15 cm] {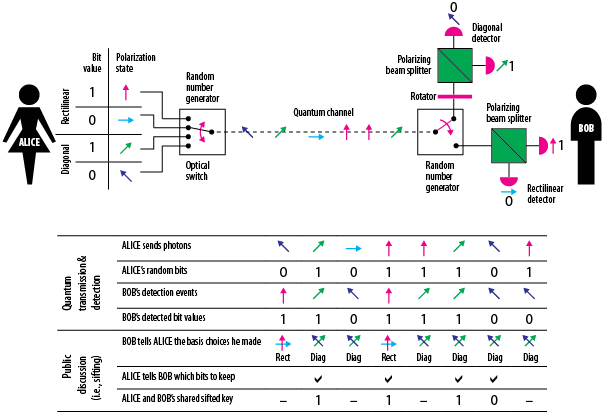}
    \caption{Quantum Cryptography Public Key Using BB84 Protocol. {\tiny (Image source: UNS Nice (France), Department of Physics)} }
    \label{fig:qkd_desc}
\end{figure}

Currently, there are many kinds of QKD protocols, and some are listed in Table \ref{tab:dvqkd}.  Discrete-Variable (DV-QKD) Protocols use photon polarization states to encode bits to generate confidential keys between sender and receiver. It also implements post-processing methods and photon counting techniques to detect the single photon to develop secret keys \cite{Bennet2020apply}. BB84 is the first QKD \cite{Scarani2009apply} protocol of this family. A basic QKD illustration is shown in Figure \ref{fig:qkd_desc}.

Typically, a single bit of information is encoded on a single photon. The bit can be stored on any basis. The two most commonly used bases are (i) vertical (V) or horizontal (H) polarization states corresponding to  ${\ket{0},\ket{1}}$ basis, and (ii)  +45° and -45° states of polarization corresponding to ${\ket{+},\ket{-}}$ basis. Both Alice and Bob use one of these two bases randomly for preparation and measurement respectively. If Bob measures in a base that is different from the one Alice used to prepare, then his answer is discarded. However, when they both use the same basis, it  results in a perfectly correlated outcome. This requires Alice and Bob to communicate about the choice of basis publicly. The QKD protocol is designed so that this public discussion does not lead to any information leakage to a potential Eavesdropper. Multiple variations of this basic method have been studied.

Continuous Variable (CV-QKD) Protocol was introduced fifteen years after the DV-QKD protocol with a different approach known as CV coding by Ralph \cite{Ralph1999apply}  to ensure better secure transmission of data. In order to implement the DV-QKD protocol a single photon source and a spotter are required. However, a standard positive intrinsic-negative (PIN) photo-diode device is used by CV-QKD and the photon detection technique applied by these two protocols are dissimilar. A coherent detection technique is adopted to replace the photon counting by CV-QKD protocol which is known as homodyne observation.  This is found to be highly fast, efficient, and economical. The first compressed state category of BB84 protocol \cite{Hillery2000apply,Cerf2001apply,Reid2000apply} used  Gaussian modulation and discrete methods. Lately, the CV-QKD practical demonstrations were performed to identify the coherent states of light by \cite{Grosshans2002apply, jouguet2013experimental}.

Distributed Phase Reference (DPR-QKD) Protocols are the family of QKD protocols that involve Differential Phase Shift (DPS)  and Coherent One-Way (COW) protocols, \cite{Gisin2004apply} that were developed recently. In both these protocols, a sequence of coherent states of fragile laser pulses is transmitted. While in DPS the pulses are modulated by keeping the intensity unchanged, in COW protocol the intensities are varied by keeping the pulses unchanged.

Quantum computing is prone to various kinds of attacks: quantum computing attacks (Beam Splitter, Photon Number Splitting) and classical computing attacks (Man-in-Middle, Denial-of-Service, Trojan, etc.). In this paper, we limit our attention to Photon Number Splitting and Denial of Service attacks on DV-QKD, CV-QKD and DPR-QKD protocols. Table \ref{tab:rlvqe} lists the attacks and countermeasures.

\begin{table}[htb]
    \centering
    \caption{QKD Protocols Attacks and Countermeasures }
    {\small
\begin{tabular}{ |p{3.5cm}|p{1.5cm}|p{4.5cm}|}
 \hline
Name of the Attack & Target & Countermeasures \\
 \hline
 Photon Number Splitting (PNS) & Source & Decoy states \cite{Hwang2003apply,zhou2012apply,Resch2005apply,Marcikic2006parameter},\\ & & SARG04 \cite{scarani2004quantum,Lucamrini2015apply,Vakhitov2001apply,Beveratos2002apply,Dusek1999apply} 
  \\
 \hline
Denial of Service & Any & BB84 \cite{Alasdair2020apply,Yuan2018apply,WenyuanWang2019apply,yu2019photonic}, \\
& & Software Defined Networks \cite{Hugues2021apply}\\
 \hline
 
 \end{tabular}}
    \label{tab:rlvqe}
\end{table}

\subsubsection {Machine learning techniques on quantum key distribution protocol }
\label{qkd}
Recently a  CV-QKD protocol with defense systems has been presented by researchers \cite{Yiyu2020apply}. The recommended technique may meritoriously recognize most known liabilities while reducing a small slice of secret keys and the transmission distance for recall values greater than 99\%, as per simulation data.  They proposed that  many properties of pulses would be transformed using a feature vector. These features are supplied to an artificial neural network (ANN) model as input for detecting and classifying the attacks. In order to train the ANN model, the authors have considered the effects of the current assault techniques that have measurable properties such as  local oscillator (LO) pulses and signals. Additionally, they have created a set of feature vectors labeled by using the dissimilar kinds of attack, as input to the ANN model. The datasets for training, testing, and performance evaluation are created by considering real-time attacks.  The proposed trained ANN model was able to identify irregular feature vectors and categorize them into dissimilar assault kinds. Subsequently, they established a general attack discovery model that is able to identify the most recognized attacks by using an accelerated computation. As and when Bob receives the secret keys, all of them are sequentially recorded into the ANN input model; at any point when abnormal data is noticed, then automatically the transmission process is aborted. Hence, Bob does not have to wait for the pending key, and the broadcast process is accomplished to verify whether the system is confronted. They have performed a simulation and revealed the trained ANN results that can automatically recognize and categorize attacks with recall and accuracy above 99\%. An interesting matter in this work is that the performance of the trained ANN model depends on the number of neurons in the hidden layer. Hence, selecting suitable values of neurons plays a vital role in real-time deployment. The proposed work marginally reduced the transmission distance and secret key rate, but nevertheless created a general defense model for most recognized attack tactics and pointedly improved the system security.

The paper  \cite{Marcine2019apply} proposes an error reconciliation strategy based on ANNs, more specifically, the Tree Parity machine (TPM). TPMs are classical cryptographic techniques to share  secure keys based on neural networks. The idea is to create a neural network with a similar structure (number of inputs, hidden layers, hidden nodes, etc.) for both the parties. The network has a single bit as output which is either +1 or -1. Both Alice and Bob start with some random initial weights and then both give the same input to the network. Whenever the outputs at both ends are the same, the training enhances weights corresponding to the fired neurons. Whenever the outputs of the networks differ, another input is chosen. Once when both the parties start getting the same outputs for multiple inputs, the two networks are synchronized and thus have roughly equal weights, which can then be used for cryptography purposes.

The concept of TPM was applied to quantum cryptography since using QKD, the two parties start with very similar bit strings with an average error of around 5-7\%. The bit strings are translated to weights of TPMs and then trained usually. Since the difference is already very low, the training can be completed much faster. Further, since Alice and Bob are not communicating parity information anymore (as is the case in conventional error reconciliation techniques such as cascade), an eavesdropper cannot get as much information out of the error reconciliation data being transferred between the two parties. Therefore, this improves both the performance and security of the protocol.

One major drawback of this approach is that it works only when the Quantum Bit Error Rate (QBER) is very low. For high QBER systems, synchronisation process by the protocol is prolonged, just like a regular TPM. Therefore, it cannot help us distinguish whether noise caused in the transmission is due to an eavesdropper or due to inherent errors in the apparatus or channel noise. That would require a scheme which does not consider the QKD process as a black box and consider the inner functioning and operations of the apparatus and the kind of noise they incur on the key exchange.

\subsubsection{Future Scope of ML in Quantum Communications}

A potential focus on quantum communication is on discovering an efficient way to transfer information in quantum channels at the nano-scale. That would include a next-generation approach of using most elegant electronics and photonic devices. These technology devices ensure the transfer of information without any losses. A promising project TOCHA \cite{tocha2019apply} by the European Commission is already in action with an objective to build a quantum particle of electromagnetic radiation such as light and photons that can be communicated with the smallest possible wastage of energy with novel topological wave guides. The current major communication challenges witnessed are relevant to both quantum-assisted classical and pure quantum communication because quantum networks generally work on the principle of disseminating single photons of light through free space or optical fiber. Hence, the new challenges are:

\begin{enumerate}
  \item Appropriate encoding schemes
  \item Quantum state generation
  \item Quantum state transmission
  \item Quantum state detection
\end{enumerate}

Further, linear optics gives hope for quantum communication challenges, where more than a single bit of quantum information is used as a carrier. These complex quantum states are capable of noise resistance in numerous configurations and allow the encoding of more data into a single photon. Additionally, in the near future from short to medium term, numerous quantum communications techniques are targeted. However, in the long term, building a Very Wide Area Network (VWAN) across continents with quantum processors is to be addressed. In the near future, QKD will probably be witnessed in showing vast distances across trusted nodes, high-altitude platform systems, test-bed networks, or satellites, as well as intra-city networks with many nodes or switchable networks, all of which would necessitate large-scale investments in infrastructure \cite{bisio2007apply}.

The engineering and operational systems are targeted at enabling high-speed electronics and optoelectronics, such as Field Programmable Gate Array/Application Specific Integrated Circuit (FPGA/ASIC)\cite{Li2021apply}, packaging, coupled photonics, and compact cryptosystems to be scaled and mass-produced in large quantities in order to deliver solutions that operate with existing communication networks. This includes combining classical and quantum encryption techniques to provide comprehensive security solutions while also expanding the market for new applications.

\section{Concluding Remarks}
\label{con}
The field of quantum computing is witnessing exponential growth in many computational applications. Employing quantum computing for efficient processing of information of massive volumes of data is the ultimate goal. This paper is a brief survey of the various applications of ML-enabled quantum computing. Machine learning is in its prime as many techniques have been used for various  computational applications. We have focused here on how machine learning techniques can be harnessed for various applications in the domain of quantum computing and communications. More specifically, the paper begins with a review of quantum computing and machine learning. It provides various ways in which these two domains have crossed paths and furthermore how the new idea of machine learning-enabled quantum computing has become a topic of global importance. 

 This article also sheds some light on the challenges associated with the implementation of these in near-term devices. The domain of AI enabled quantum computing is still nascent with a lot of potential and many research opportunities.

In a nutshell, machine learning-enabled quantum computing is a new and fast-growing area, yet   many challenges remain before it can be at par with classical ML. Our aim is to provide a basic understanding, trigger interest, and supply ample references for further in-depth studies, to both beginners and experts in the various domains that involve both QC and ML.

\section*{Acknowledgments}

We would like to acknowledge the fruitful discussion with Ritajit Majumdar, a former Senior Research Fellow at the Indian Statistical Institute and presently a Research Scientist at IBM India Research Lab.

\appendix

\section{Appendix}

\subsection*{Quantum Circuit}
A quantum circuit serves as the schematic depiction of a quantum algorithm or quantum program. Every line within the quantum circuit is denoted by a qubit, and the operations, specifically quantum gates, are depicted using distinct blocks placed along the line \cite{barenco1995elementary}. Table 5 provides a summary of frequently used logical quantum gates. 



\begin{figure}[hbt]
    \centering
     \caption*{Table 5: Matrix Representation of Quantum Gates\cite{basu2022qer}}
    \includegraphics[height=16 cm, width= 15 cm] {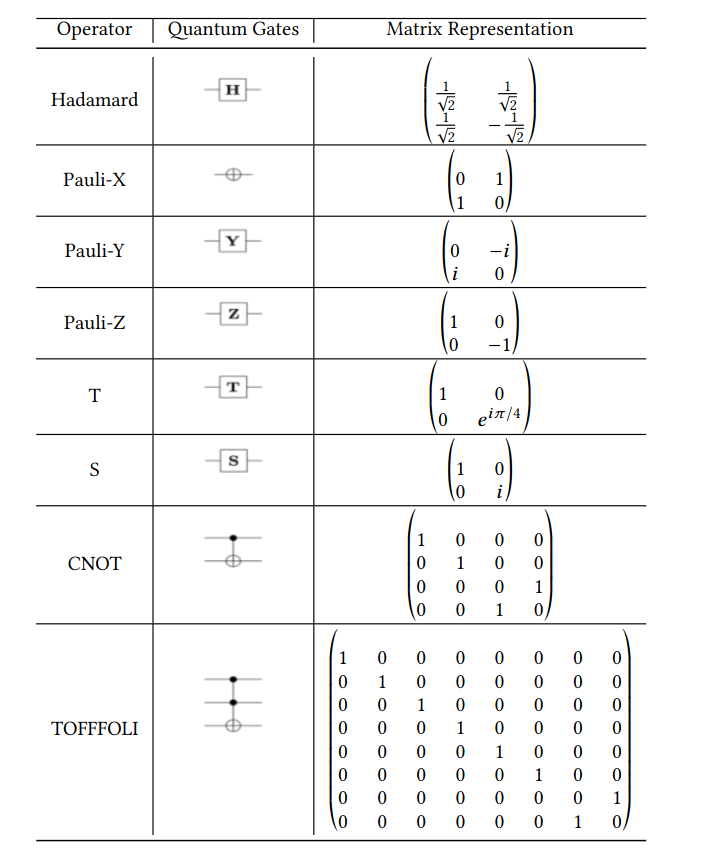}
   
    \label{fig:appen}
\end{figure}

\newpage
\bibliographystyle{unsrt}
\bibliography{main}

\end{document}